\documentclass[12pt]{iopart}
\usepackage{ucs}
\usepackage{multirow}
\usepackage{longtable}
\usepackage{ragged2e}
\usepackage{graphicx}% Include figure files
\usepackage{dcolumn}% Align table columns on decimal point
\usepackage{hyperref}% add hypertext capabilities
\usepackage{blindtext}
\usepackage{cite}
\usepackage{xcolor}
\usepackage[normalem]{ulem} % to strike out text

\begin{document}
\title[Improving the spectroscopic knowledge of neutral Neodymium]{Improving the spectroscopic knowledge of neutral Neodymium}

\author{Gohar \; Hovhannesyan$^1$, Maxence \; Lepers$^1$}

\address{$^1$ Laboratoire Interdisciplinaire Carnot de Bourgogne, UMR 6303 CNRS Univ. Bourgogne Franche-Comt\'{e}, 9 Avenue Alain Savary, BP 47 870, F-21078 DIJON Cedex FRANCE}

\ead{gohar.hovhannesyan@u-bourgogne.fr}

\begin{abstract}
Laser cooling and trapping of lanthanides has opened the possibility to carry out new experiments with ultracold dipolar gases, for example for quantum simulation of solid state physics. To identify new suitable candidates for laser-cooling, it is important to have a precise spectroscopic knowledge of the atom under consideration. Along this direction, we present here a detailed modeling of the energy levels of neutral neodymium (Nd), an element belonging to the left part of the lanthanide row, which has not yet been considered for laser-cooling. Using the semi-empirical method implemented in the Cowan suite of codes, we are in particular able to interpret more than 200 experimental levels of the NIST database belonging to both parities. The optimal set of atomic parameters obtained after the least-square fitting step can serve to calculate radiative transition probabilities in the future.
%Laser cooling has found wide application in science and technology. The success of laser cooling and trapping of lanthanides makes it possible to carry out new experiments with ultracold dipolar gases, for example for quantum simulation of solid state physics. In this article we cautiously study the atomic energies and transition dipole moments (TDMs) for neutral Neodymium.  To obtain a detailed knowledge of its spectroscopy, we use COWAN suite of codes, based on a semi-empirical method. In this work, we calculate spectroscopic data for six configurations of neutral Nd, three for each parity.

\end{abstract}
%\keywords{lanthanides, Neodymium, spectroscopic calculations}

\accepted{}
\maketitle
\ioptwocol
\section{Introduction}
In the field of ultracold atomic and molecular matter, quantum gases composed of particles with a strong intrinsic permanent dipole moment, called dipolar gases, have attracted great interest in the last few years because they can be controlled by external electric field or magnetic fields.
Through long-range and anisotropic interactions between particles, dipolar gases enable the production and study of highly correlated quantum matter, which is critical for quantum information or for modeling many-body or condensed matter physics \cite{baranov2008, bloch2008, bloch2012}.

Among the different families of systems, open-shell atoms have a permanent magnetic dipole moment that is determined by their total electronic angular momentum. In the context of ultracold matter, important achievements were the first Bose-Einstein condensates of highly magnetic atoms obtained with chromium \cite{griesmaier2005, beaufils2008}. Later, much attention began to be attracted to the lanthanides, a series of 15 elements with atomic numbers $Z$ = 57--71, from lanthanum (La) through lutetium (Lu). Lanthanides, along with the chemically similar elements scandium and yttrium, are often collectively known as the rare earth elements.
Lanthanide atoms open up new possibilities for interactions, not only because of their large ground state magnetic dipole moments, but also because of the large number of optical transitions with widely varying properties that provide a better controllability, or because of pairs of quasi-degenerate metastable levels, allowing the production of an electric and magnetic dipolar gases \cite{lepers2018}. Finally, the lanthanides have the great advantage of having fermionic and/or bosonic stable isotopes.

These distinctive properties are primarily due to a unique electronic structure: the so-called submerged f-shell configuration. Most lanthanides have a completely filled 6s shell and a partially filled inner 4f shell. Moreover, among the elements with the largest atomic numbers, many share a common set of properties and often have similar transitions at the same wavelengths \cite{norcia2021, chomaz2022}.
So far, laser cooling has been demonstrated for elements belonging to the right part of the lanthanide row, namely erbium \cite{frisch2012, seo2020efficient, ban2005laser, berglund2008narrow},
dysprosium \cite{lu2010trapping, lunden2020enhancing, lu2011spectroscopy, dreon2017optical}, holmium \cite{miao2014magneto}, thulium \cite{sukachev2010magneto, vishnyakova2014two} and europium \cite{inoue2018magneto}, as well
as in erbium–dysprosium mixtures \cite{ilzhofer2018two}.
%Ultracold erbium and dysprosium atoms were created in a magneto-optical trap, and later, noncondensed ultracold gases of thulium and holmium were produced.

These achievements open the question of identifying new suitable species for laser cooling, especially in the left part of the lanthanide series. Among them, we notice that, cerium (Ce, $Z = 58$) has the ground configuration 4f\,5d\,6s$^2$, which makes this element \textit{a priori} not convenient for such experiments. On the other hand, when we go to the middle of the series, we have radioactive promethium (Pm, $Z = 61$), after which the spectrum of the elements becomes more and more dense, starting with samarium (Sm, $Z = 62$), making these elements not favourable for possible laser cooling studies.
Therefore, neodymium (Nd, $Z = 60$) and praseodymium (Pr, $Z = 59$) represent the most promising energy spectrum for the formation of a dipolar gas. Their lowest configurations are very close in energy, namely 4f$^n$ 6s$^2$, 4f$^n$\,5d\,6s, 4f$^{n-1}$\,5d\,6s$^2$ and 4f$^{n-1}$\,5d$^2$\,6s, where $n = 3$ for Pr and $n = 4$ for Nd. The lowest levels of 4f$^n$\,6s$^2$ and 4f$^{n-1}$\,5d\,6s$^2$ mainly have a spin equal to $S = n/2$, denoting that laser-cooling transitions may be chosen among these configurations. Meanwhile, the lowest levels of 4f$^n$\,5d\,6s and 4f$^{n-1}$\,5d$^2$\,6s configurations are mainly characterized by a spin $S = n/2 + 1$, which makes the decay by spontaneous emission toward levels of 4f$^n$\,6s$^2$ and 4f$^{n-1}$\,5d\,6s$^2$ rather unlikely. The 4f$^n$\,5d\,6s and 4f$^{n-1}$\,5d$^2$\,6s configurations also have levels that are very close in energy, and that can be significantly mixed to induce an electric dipole moment \cite{lepers2018}. Moreover Nd represents the great advantage of having bosonic and fermionic stable isotopes, while Pr has only one bosonic stable isotope.
%Their four lowest electronic configurations are very close in energy. The lowest levels have an electronic spin equal to $S = n/2$.

\begin{figure}
    \centering
    \includegraphics[scale=2.35]{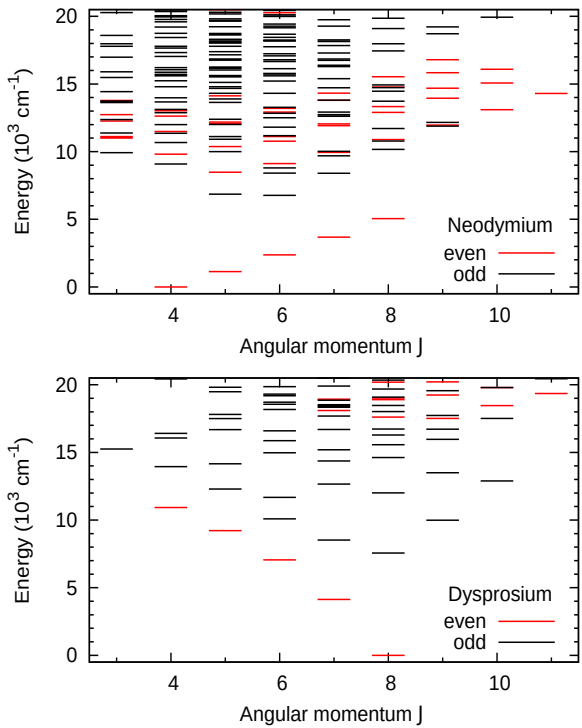}

    \caption{Energy diagrams as functions of the electronic angular momentum $J$ and sorted by electronic parity for neodymium (Nd, top panel) and dysprosium (Dy, bottom panel).}
    \label{fig:Nd_Dy_lev}
\end{figure}

In order to find possible laser-cooling transitions for neutral Nd, it is essential to carefully model the spectrum, \textit{i.e.}~energies and transition dipole moments (TDMs). In this work, as a first step, we carefully study the Nd energy levels. Particular attention is paid to accurately describing configuration-interaction (CI) mixing, to which TDMs are very sensitive, especially those that lead to weak transitions, which play an important role in this design. Since Nd belongs to the left part of the lanthanide row, it presents a high density of levels in the range 8000-15000~cm$^{-1}$ in contrast with Dy (see figure \ref{fig:Nd_Dy_lev}).
To calculate energies, we use the combination of \textit{ab initio} and least-square fitting techniques implemented in the Cowan codes \cite{cowan1981, kramida2019}. We include the three lowest configurations of each parity which allows us to interpret more than 200 energy levels given in the NIST ASD database  \cite{NIST_ASD}. The main technical difficulty of this work comes from the least-squares fitting of close energy levels, because we need to determine to which experimental counterparts each computed level should converge.

%Evidently, the success of projects aimed at laser cooling of neutral lanthanides requires a thorough study of the spectroscopic data of the corresponding elements.

The article is organized as follows: in section \ref{sec:methodology} we describe the general methodology of our spectroscopic calculation. Then section \ref{sec:calcultions_results} is devoted to the calculation of neutral Nd. In that section we also present the results of the calculations divided in several steps and we conclude the work in section \ref{sec:conclusion}.

\section{Methodology}
\label{sec:methodology}

The calculations of the neutral Nd spectrum are performed with the semi-empirical technique provided by Robert Cowan's atomic-structure suite of codes, for which we used both the McGuinness \cite{mcguinness_cowan} and the Kramida \cite{kramida2019} versions, and whose theoretical background is presented in \cite{cowan1981}. In the present section, we briefly review the principles of those calculations.

As a first step, \textit{ab initio} single-electron radial wave functions $P_{n\ell}$ for all the subshells $n\ell$ of the considered configurations are computed with the relativistic Hartree-Fock (HFR) method. The principal output, for each configuration, consists of energy parameters, such as center-of-gravity configuration energies $E_\mathrm{av}$, direct $F^k$ and exchange $G^k$ electrostatic integrals, or spin-orbit integrals $\zeta_{n\ell}$, that are the building blocks of the atomic Hamiltonian and are required to calculated the energy levels. For each couple of configurations, the wave functions $P_{nl}$ serve also to calculate the CI parameters $R^k$.

In a second step, the program sets up energy matrices for each possible value of total angular momentum $J$, diagonalizes each matrix to get eigenvalues and eigenvectors. It is possible to calculate Land{\'e} $g$-factors, as well as electric-dipole (E1), electric quadrupole (E2) and magnetic-dipole (M1) radiation spectra with wavelengths, oscillator strengths, radiative transition probabilities and radiative lifetimes.
It is important to emphasize that the basis functions used by the codes are the numerical functions obtained after the HFR calculation for each configuration, which are then combined appropriately to describe the atom in the desired angular momentum coupling scheme, i.e., \textit{LS}, \textit{jj} or others.

In \textit{LS} or Russell-Saunders coupling conditions the electrostatic interactions between electrons are much stronger than the interaction between the spin of an electron and its own orbital motion. In this case, an atomic level is in particular characterized by its total orbital and its total spin quantum numbers,  $L$ and $S$. With increasing $Z$, the spin-orbit interactions become increasingly more important. When these interactions become much stronger than the Coulombic terms, the coupling conditions approach pure $j j$ coupling case. In the present study, atomic levels are usually well represented in intermediate coupling, \textit{i.e.}~their eigenvectors are sums of basis states written in \textit{LS} coupling.

\begin{figure}
    \centering
    \includegraphics{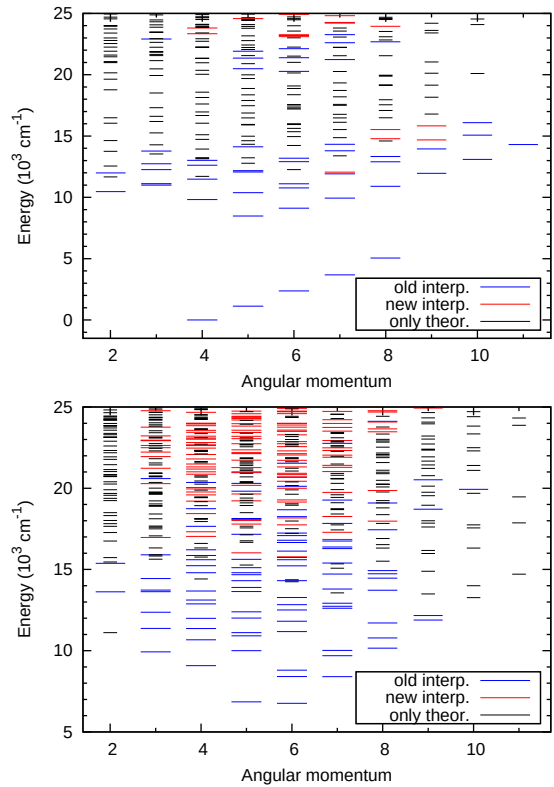}

    \caption{Experimental (\textit{blue}), newly interpreted (\textit{red}) and newly predicted (only theoretical) (\textit{black, short}) energy levels of even parity (top) and odd parity (bottom) configurations of neutral Nd as functions of the electronic angular momentum $J$. Plots are limited to energy values of 25000~cm$^{-1}$.}
    \label{fig:Nd_new_lev}
\end{figure}

When higher accuracy is desired, in a third step, radial energy parameters are treated as adjustable parameters of a least-square fitting calculation. This is done in order to find the best possible agreement between the Hamiltonian eigenvalues and the experimental energies. As experimental energies we use the data published in the NIST database \cite{NIST_ASD}.
The accuracy of the fit is measured by means of the standard deviation:
\begin{equation}
  s = \left[ \frac{ \sum_{i=1}^{N_\mathrm{lev}} 
    (E_{\mathrm{th},i}-E_{\mathrm{exp},i})^2 }
    { N_\mathrm{lev}-N_\mathrm{par} }  \right]^{\frac{1}{2}},
  \label{eq_stdev}
\end{equation}
where $E_{\mathrm{exp},i}$ are the observed energy values and $E_{\mathrm{th},i}$ are the computed eigenvalues, $N_\mathrm{lev}$ is the number of levels being fitted and $N_\mathrm{par}$ is the number of adjustable parameters (or parameter groups) involved in the fit \cite{cowan1981}.

In an attempt to improve the quality of the fit, a variety of ``effective-operator" parameters, called $\alpha$, $\beta$ and $\gamma$, also known as the effective electrostatic parameters or Trees parameters, and {}``illegal-k'' $F^k$, $G^k$ have been introduced, representing corrections to both the electrostatic and the magnetic single-configuration effects \cite{cowan1981}. {}``Illegal-k'' means that these are the values of $k$ for which $k + \ell + \ell'$ is odd,
where $\ell$ and $\ell'$ are the orbital angular momenta of the electrons coupled by the effective operator; for example ($\ell$,~$\ell'$)=(3, 2) for (4f, 5d) electrons. These effective parameters, unlike other parameters, can not be calculated \textit{ab initio}, but are there to compensate the absence of electronic configurations not included in the model.
Due to the lack of HFR estimates, the initial values of the effective parameters are obtained from comparisons with similar spectra.

To make some comparisons between different elements and ionization stages, one often defines the scaling factor (SF) $f_X = X_\mathrm{fit} / X_\mathrm{HFR}$ between the fitted and the HFR value of a given parameter $X$.
During the fitting procedure, it is sometimes convenient to be able to link several parameters together in such a way that their SFs remain identical throughout the calculation; such groups of constrained parameters are characterized by the same $r_n$ value in tables \ref{tab:Nd_SF_even} - \ref{tab:Nd_SF_CI}. The word ``fixed'' means that the corresponding parameters are not adjusted.

%For neutral Neodymium (Nd I), the experimental energies are published in the NIST database \cite{NIST_ASD}, constructed from the report of Martin \textit{et al} \cite{martin1978}.

%In an attempt to improve the quality of the fit (and therefore, the accuracy of the resulting eigenvectors), a variety of ``effective-operator" parameters, called $\alpha$, $\beta$ and $\gamma$ and {}``illegal''-k $F^k$, $G^k$ have been introduced, representing corrections to both the electrostatic and the magnetic single-configuration effects \cite{cowan1981}. {}``Illegal''-k means that these are the values of $k$ for which $k + \ell + \ell'$ is odd. These effective parameters, unlike other parameters, can not be calculated \textit{ab initio}.

%The general methodology for our fitting calculations is as follows: (a) fitting the parameters with an \textit{ab initio} values while effective parameters are forced to be zero; (b) fixing the parameters resulting from step (a) and fitting the effective parameters; (c) using the final values of (b), fitting all the parameters together.

\section{Results for Nd}
\label{sec:calcultions_results}

This section, dedicated to our results for Nd, is divided as follows. In subsection \ref{sub_calc} we present the different steps of our least-square fitting calculation, discussing especially the configurations included, the number of fitting parameter groups, and the resulting standard deviations. In the next subsections, we describe in more details our results, where: (i) we include in the fit levels of the NIST database that are interpreted, see subsection~\ref{sub_nist-interp}; and (ii) we include levels that are not interpreted, see subsection~\ref{sub_new_interp}.

\subsection{Description of calculations}
\label{sub_calc}

%\begin{figure}
%    \centering
 %   \includegraphics{Nd-I_lev_new_interp_even.eps}
%    \includegraphics{Nd-I_lev_new_interp_odd.eps}
  %  \hspace{.7cm}

%    \caption{Experimental (\textit{blue}), newly interpreted (\textit{red}) and newly identified (only theoretical) (\textit{black, short}) energy levels of even configurations of neutral Nd as functions of the electronic angular momentum $J$. Plots are limited to energy values of 25000~cm$^{-1}$.}
%    \label{fig:Nd_new_lev_odd}
%\end{figure}

The calculations were performed with three configurations in each parity, namely:
\begin{itemize}
\item 4f$^4$\,6s$^2$, 4f$^4$\,5d\,6s, 4f$^3$\,5d\,6s\,6p for the even parity;
\item 4f$^4$\,6s\,6p, 4f$^3$\,5d\,6s$^2$, 4f$^3$\,5d$^2$\,6s for the odd parity.
\end{itemize}
For both parities, we use values from the NIST database as reference energy levels \cite{NIST_ASD}. The primary source of data on neutral Nd levels in the NIST database is Martin \textit{et al.} \cite{martin1978}.

\begin{figure}
    \centering
    \includegraphics[scale=0.45]{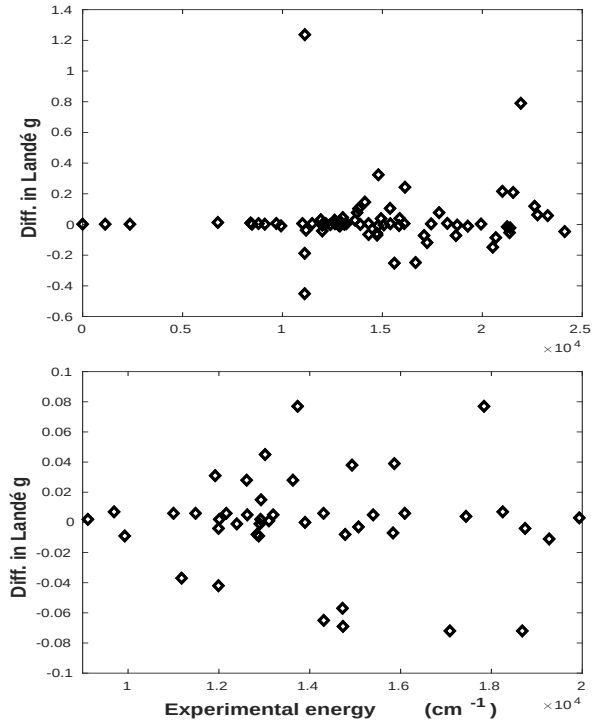}
    \caption{Differences between calculated and experimental Land\'e $g$-factors for energy levels with an experimentally known $g$-factor. The picture on the bottom is an enlarged version to show the differences in detail. The largest difference at 11129 cm$^{-1}$ (top panel) is out of the y-scale on the bottom panel. Energy levels are in cm$^{-1}$.}
    \label{fig:Lande_g_diff}
\end{figure}

\begin{table*}
\centering
\caption{\label{tab:Nd_diff_case1} Differences between NIST database and theoretical results for energies, Landé-g factors and dominant \textit{LS} terms with the percentage of the theoretical one. Case 1: when the configurations are the same, but there are differences in terms. All energy levels are in cm$^{-1}$.
%Here we use the following abbreviations: conf.=configuration, exp.=experimental data taken from the NIST database and dom. term=dominant eigenvalue term with its percentage.
}
\lineup
%\begin{indented}
%\item[]
\begin{tabular} {@{}lllllllll}
\\
\br
& & \multicolumn{2}{c}{Energy} & \multicolumn{2}{c}{Land\'e g } & \multicolumn{3}{c}{Dominant term } \\ \cline{3-9} \\
Configuration & J  & Theory & Exp. & Theory & Exp. & \multicolumn{2}{c}{Theory} & Exp. \\
\cline{1-2} \cline{3-4} \cline{5-6} \cline{7-9}\\
%new
B-d6sp & 7 & 24187 & 24218 & 1.095 & 0.870 & 58\% & $^7$K$\phantom{A}$ & $^5$M \\
B-d2s & 4 & 15457 & 15600 & 0.704 & 0.630 & 15\% & $^5$I$\phantom{A}$ & $^7$G \\
A-6sp & 4 & 20273 & 20361 & 0.957 & 0.735 & 27\% & $^5$H$\phantom{A}$ & $^5$I \\
A-6sp & 5 & 20271 & 20301 & 1.169 & 0.775 & 35\% & $^7$G$\phantom{A}$ & $^5$K \\
A-6sp & 5 & 21015 & 21005 & 1.176 & 0.960 & 28\% & $^7$F$\phantom{A}$ & $^5$I \\
B-d2s & 6 & 15522 & 15598 & 0.958 & 1.210 & 31\% & $^7$K$\phantom{A}$ & $^7$H \\
B-d2s & 6 & 18535 & 18679 & 1.008 & 1.080 & 17\% & $^3$K$\phantom{A}$ & $^7$I \\
B-ds2 & 6 & 20112 & 20119 & 1.039 & 1.015 & 21\% & $^5$H$\phantom{A}$ & $^3$K \\
B-d2s & 7 & 16633 & 16747 & 1.059 & 1.265 & 21\% & $^7$K$\phantom{A}$ & $^7$H \\

%B-ds2 & 3 & 20649 & 20595 & 1.147 & 0.910 & 17\% & $^5$P$\phantom{A}$ & $^3$G \\
%B-ds2 & 5 & 19934 & 19816 & 1.064 & 1.110 & 11\% & $^3$H$\phantom{A}$ & $^5$H \\
%B-ds2 & 6 & 13078 & 13277 & 1.058 & 0.990 & 22\% & $^3$I$\phantom{A}$ & $^3$K \\
%B-d2s & 6 & 15509 & 15598 & 0.957 & 1.210 & 29\% & $^7$K$\phantom{A}$ & $^7$H \\
%B-d2s & 6 & 16133 & 16128 & 1.191 & 0.950 & 38\% & $^7$H$\phantom{A}$ & $^5$K \\
%B-d2s & 6 & 17270 & 17237 & 1.040 & 1.190 & 10\% & $^7$K$\phantom{A}$ & $^7$H \\
%B-d2s & 7 & 19222 & 19271 & 1.208 & 1.260 & 17\% & $^7$I$\phantom{A}$ & $^7$G \\
\br
\end{tabular}
%\end{indented}
%\end{ruledtabular}
\end{table*}

Since it belongs to the left part of the lanthanide row of the Periodic Table of the Elements, Nd possesses a dense spectrum, which makes it difficult to identify the levels. In order to overcome this issue, we have divided the calculation into steps.
As a first step, for even parity, the configurations 4f$^4$\,6s$^2$ and 4f$^4$\,5d\,6s were considered together, and the calculations for the configuration 4f$^3$\,5d\,6s\,6p were carried out separately. For the first group when we have included 42 experimental levels and the fitting is done with 11 groups of parameters, the standard deviation is 91~cm$^{-1}$. For the configuration 4f$^3$\,5d\,6s\,6p the calculations were done with 10 groups of parameters. When 14 interpreted experimental levels are included the standard deviation is 101~cm$^{-1}$.

\begin{table*}
\caption{\label{tab:Nd_diff_case2} Differences between NIST database and theoretical results for energies, Landé-g factors and dominant LS terms, with the percentage of the theoretical one. Case 2: when there is a good match in energy levels, but the configurations are different. All energy levels are in cm$^{-1}$.
%Here we use the following abbreviations: exp.=experimental data taken from the NIST database and dom. term=dominant eigenvalue term with its percentage.
}
\lineup
\centering
%\begin{indented}
%\item[]
\begin{tabular}{@{}llllllllrl}
\\
\br
J & \multicolumn{2}{c}{Energy } & \multicolumn{2}{c}{Configuration } & \multicolumn{2}{c}{Land\'e g } & \multicolumn{3}{c}{Dominant term }\\ \mr \\
 & Theory & Exp. & Theory & Exp. & Theory & Exp. & \multicolumn{2}{l}{Theory} & Exp. \\
\cline{2-3} \cline{4-5} \cline{6-7} \cline{8-10} \\
4 & 14716 & 14802 & B-d2s & A-6sp  & 0.443 & 0.825 & 80\% & $^7$K$\phantom{A}$ & $^5$H \\
4 & 15898 & 15863 & B-ds2 & A-6sp & 1.059 & 1.020 & 62\% & $^3$G$\phantom{A}$ & $^7$H \\
4 & 16293 & 16210 & B-d2s & B-ds2 & 0.771 & 1.055 & 66\% & $^7$I$\phantom{A}$ & $^3$G \\
4 & 18701 & 18741 & B-d2s & B-ds2 & 0.926 & 0.930 & 8\% & $^5$I$\phantom{A}$ & $^5$H \\
5 & 15049 & 14797 & B-ds2 & B-d2s & 1.084 & 0.760 & 37\% & $^3$H$\phantom{A}$ & $^5$K \\
5 & 15215 & 15114 & B-d2s & B-ds2 & 0.872 & 1.110 & 27\% & $^7$K$\phantom{A}$ & $^3$H \\
%7 & 16272 & 16283 & B-ds2 & A-6sp & 1.199 & 1.110 & 41\% & $^5$H$\phantom{A}$ & $^5$H \\
%7 & 16302 & 16388 & A-6sp & B-ds2 & 1.107 & 1.175 & 86\% & $^7$K$\phantom{A}$ & $^7$K \\
7 & 22752 & 22761 & B-d2s & A-6sp & 1.098 & 1.035 & 14\% & $^3$K$\phantom{A}$ & $^5$K \\
8 & 24148 & 24121 & B-d2s & A-6sp & 1.089 & 1.135 & 12\% & $^5$L$\phantom{A}$ & $^5$K \\
9 & 20594 & 20523 & B-d2s & A-6sp & 1.082 & 1.230 & 42\% & $^5$M$\phantom{A}$ & $^7$I \\
%5 & 20988 & 21005 & B-d2s & A-6sp & 0.927 & 0.960 & 34\% & $^5$I$\phantom{A}$ & $^5$I \\
%6 & 16487 & 16658 & B-d2s & A-6sp & 0.802 & 1.050 & 54\% & $^5$L$\phantom{A}$ & $^7$I \\
%6 & 16799 & 16797 & A-6sp & B-d2s & 1.141 & 0.955 & 91\% & $^7$I$\phantom{A}$ & $^7$K \\
%6 & 20655 & 20673 & A-6sp & B-ds2 & 1.340 & 1.185 & 72\% & $^7$G$\phantom{A}$ & $^5$H \\
\br
\end{tabular}
%\end{indented}
\end{table*}

After the calculation, the optimal values of the energy parameters were determined. In the next step, these two groups were combined together, and the optimal parameters of the individual calculations were taken as an initial set for the combined calculation.
In this step 54 interpreted experimental levels are included for three even parity configurations and the fitting is done with 12 groups of free parameters. The standard deviation for this combined calculation is 89~cm$^{-1}$. The latter results are discussed in more details in subsection \ref{sub_nist-interp}.

We followed a similar method for odd parity configurations. We have treated separately the configurations 4f$^3$\,5d\,6s$^2$ and 4f$^3$\,5d$^2$\,6s on one hand, and 4f$^4$\,6s\,6p on the other hand. For the first group of odd parity configurations the calculation is done with 11 parameter groups and 79 experimental levels are included. After the final calculation the standard deviation is 94~cm$^{-1}$. For configuration 4f$^4$\,6s\,6p we have 19 experimental levels included and 10 parameter groups. For this configuration standard deviation is 160~cm$^{-1}$.
When these two separate analyzes have been completed, we treated these three configurations together. The final least square fitting is done with 15 parameter groups and there are 96 levels included. Standard deviation in this case is 111~cm$^{-1}$. Again, the latter results are discussed in more details in subsection \ref{sub_nist-interp}.

\begin{table*}
\centering
\caption{\label{tab:Nd_diff_case3} Differences between NIST database and theoretical results for energies, Landé-g factors and dominant LS terms, with the percentage of the theoretical one. Case 3: when the configurations are different, but among the other components of the level eigenvectors, there is one whose configuration or term make identification possible (see the last three columns). All energy levels are in cm$^{-1}$.
%Here we use the following abbreviations: th.=theory, those correspond to our calculation results, exp.=experimental data taken from the NIST database, conf.=configuration and dom. term=dominant eigenvalue term with its percentage.
}
\lineup
%\begin{indented}
%\item[]
\begin{tabular} {@{}lllllllrrlrr}
\\
\br
J & \multicolumn{2}{c}{Energy} & \multicolumn{2}{c}{Configuration} & \multicolumn{2}{c}{Land\'e g}  & \multicolumn{3}{c}{Dominant term} & \multicolumn{2}{c}{Other component} \\ \cline{1-12} \\
 & Theory & Exp. & Theory & Exp. & Theory & Exp. & \multicolumn{2}{c}{Theory} & Exp. & \\
\cline{2-3} \cline{4-5} \cline{6-7} \cline{8-10} \\
3 & 15886 & 15899 & A-6sp & B-d2s & 0.737 & 0.600$\phantom{A}$ & 48\% & $^7$H$\phantom{A}$ & $^5$H & B-d2s$\phantom{A}$ 18\% & $^7$I \\
3 & 20600 & 20595 & B-d2s & B-ds2 & 1.037 & 0.910$\phantom{A}$ & 11\% & $^5$H$\phantom{A}$ & $^3$G & B-ds2$\phantom{A1}$ 9\% & $^5$P \\
5 & 19912 & 19816 & B-d2s & B-ds2 & 1.016 & 1.110$\phantom{A}$ & 8\% & $^7$H$\phantom{A}$ & $^5$H & B-ds2$\phantom{A1}$ 7\% & $^3$H \\
6 & 14270 & 14308 & B-d2s & B-ds2 & 1.041 & 1.106$\phantom{A}$ & 30\% & $^7$I$\phantom{A}$ & $^5$H & B-ds2$\phantom{A}$ 16\% & $^5$H \\
6 & 20690 & 20673 & B-d2s & B-ds2 & 1.099 & 1.185$\phantom{A}$ & 16\% & $^5$I$\phantom{A}$ & $^5$H & B-ds2$\phantom{A1}$ 7\% & $^3$I \\
6 & 21548 & 21543 & B-d2s & A-6sp & 1.109 & 0.900$\phantom{A}$ & 7\% & $^5$H$\phantom{A}$ &$^5$K & A-6sp$\phantom{A1}$ 6\% & $^5$I \\
7 & 19192 & 19271 & A-6sp & B-d2s & 1.249 & 1.260$\phantom{A}$ & 48\% & $^7$H$\phantom{A}$ & $^7$G & B-d2s$\phantom{A}$ 11\% & $^7$I \\
9 & 25649 & 25519 & B-d2s & A-6sp & 1.205 & 1.220$\phantom{A}$ & 17\% & $^5$K$\phantom{A}$ & $^5$K & A-6sp$\phantom{A}$ 16\% & $^5$K \\
%|good result| 5 & 15756 & 15626 & A-6sp & A-6sp & 0.982 & 0.850 & $^7$I & 19\% & $^7$I & A-6sp $^7$I & 16\% \\
%|good result| 5 & 17173 & 17163 & A-6sp & A-6sp & 1.105 & 1.135 & $^7$H & 57\% & - & A-6sp $^7$H & 20\% \\
%|case 1| 5 & 20271 & 20301 & A-6sp & A-6sp & 1.169 & 0.775 & $^7$G & 35\% & $^5$K & A-6sp $^5$I & 14\% \\
%|case 1| 5 & 21015 & 21005 & A-6sp & A-6sp & 1.176 & 0.960 & $^7$F & 28\% & $^5$I & A-6sp $^5$H & 5\% \\
%|case 1| 6 & 20112 & 20119 & B-ds2 & B-ds2 & 1.039 & 1.015 & $^5$H & 21\% & $^3$K & B-ds2 $^5$H & 27\% \\
%|case 2| 7 & 16272 & 16283 & B-ds2 & A-6sp & 1.199 & 1.110 & $^5$H & 41\% & $^5$H & A-6sp $^7$K & 5\% \\
%|case 2| 7 & 16302 & 16388 & A-6sp & B-ds2 & 1.107 & 1.175 & $^7$K & 86\% & $^7$K & B-ds2 $^5$H & 2\% \\
%|case 2| 8 & 24148 & 24121 & B-d2s & A-6sp & 1.089 & 1.135 & $^5$L & 12\% & $^5$K & A-6sp $^5$K & 12\% \\

\br
\end{tabular}
%\end{indented}
%\end{ruledtabular}
\end{table*}

In what follows we will use the following abbreviations for even parity configurations: 4f$^4$\,6s$^2$ = A-6s2, 4f$^4$\,5d\,6s = A-ds and 4f$^3$\,5d\,6s\,6p = B-d6sp and for odd parity configurations: 4f$^4$\,6s\,6p = A-6sp, 4f$^3$\,5d\,6s$^2$ = B-ds2 and 4f$^3$\,5d$^2$\,6s = B-d2s.

%\section{Results}
%\label{sec:results}
\subsection{NIST interpreted levels}
\label{sub_nist-interp}

In the NIST database, some of the Nd levels are well interpreted:
detailed information are given, such as Land\'e g-factors, dominant configurations, terms, etc. To distinguish these levels from other levels present in the NIST database, we refer to them as ``NIST interpreted'' levels. This subsection is devoted to the calculation when only the interpreted experimental levels are included in the fitting process.

%On the other hand, there are many levels in the database that have not been interpreted. We call these levels non-interpreted (additional) levels. The possible identification and inclusion of these levels in the calculation is discussed in the next subsection.

As stated before the dense spectrum of neutral neodymium makes it difficult to identify the levels. This is especially true for levels of $J=4$, 5 and 6. For most levels, the matching between theory and the NIST database is quite good. However, we noticed differences which can be divided into three groups:
\begin{description}
\item \textit{Case 1}: when the configurations are the same, but there are differences in the leading terms (see table \ref{tab:Nd_diff_case1}).
\item \textit{Case 2}: when there is a good match in energy levels, but the configurations are different (see table \ref{tab:Nd_diff_case2}).
\item \textit{Case 3}: when the configurations are different, but in the second or third component of the level eigenvector, the configuration and/or the term is the same as in the experimental leading term, which makes the identification possible (see table \ref{tab:Nd_diff_case3}).
%\blindtext[1]
\end{description}
Except the first level of table \ref{tab:Nd_diff_case1}, those three tables only contain levels of odd parity, mostly with intermediate angular momenta $J=4$ to 6, for which the energy spectrum is the densest. Their leading term have a low percentage (mostly below 50~\%), which means that the leading term coming out of calculations can be sensitive to the radial parameters. The corresponding optimal radial parameters and their SFs are given in the supplementary material.

%\begin{table}
%\caption{\label{tab:Nd_even} Standard deviation evolution of every step of the calculations for even parity of neutral neodymium.}
%\begin{ruledtabular}
%\begin{tabular}{ccc}
%\hline
%  & 4f$^4$6s$^2$ $+$ 4f$^4$5d6s & 4f$^3$5d6s6p \\ \cline{2-3}
% Cond. & 41 exp. lev., 11 param. & 14 exp. lev., 10 param. \\
%\hline
% 1st step & 17.93 & $9.424 \times 10^{-7}$ \\
% 2nd step & 11.92 & $2.330 \times 10^{-5}$ \\
% Int. lev. &  2.13 & $7.187 \times 10^{-8}$ \\
% Add. lev. &  2.13 & $7.187 \times 10^{-8}$

\subsection{Newly interpreted levels}
\label{sub_new_interp}

After successfully performing the calculation for six Nd configurations with NIST-interpreted levels and finding the optimal parameters for each configuration, we proceeded to include in the fit levels that are present in the database but are not interpreted. We were able to identify 25 levels for even-parity configurations and over 200 levels for odd-parity configurations (see figure \ref{fig:Nd_new_lev}). The inclusion of these new interpreted levels produced the following results:
for even parity, with 83 levels included and 12 parameter groups, the standard deviation is 90~cm$^{-1}$, and for even parity, with 298 levels included and 15 parameter groups, the standard deviation drops to 74~cm$^{-1}$.

Figure \ref{fig:Nd_new_lev} shows the energies of even and odd configurations as functions of the angular momentum $J$. Note that unlike figure \ref{fig:Nd_Dy_lev}, figure \ref{fig:Nd_new_lev} has one panel for each parity. The blue lines show the experimental energy of interpreted levels present in the NIST database, red lines correspond to the experimental energies of levels that are present in the database but have not been interpreted in detail. Finally, black short lines correspond to newly predicted levels, indicating that their energies are purely theoretical. We see that the latter are numerous and that they are located among experimental levels. In the even parity, there are no experimental levels between approximately 16000 and 20000~cm$^{-1}$, corresponding respectively to the highest interpreted levels of the 4f$^4$\,5d\,6s configuration and the lowest ones of the 4f$^3$\,5d\,6s\,6p configuration. In the odd parity, the density of levels is even larger. For extreme values of $J$, the predicted levels are significantly more present than experimental ones. This trend is not visible for intermediate values $J=4$--7 where more experimental levels were observed spectroscopically.

\begin{table*}
\centering
\caption{ \label{tab:Nd_SF_even} Parameter names, constraints, fitted values and scaling factors ($f_X = X_\mathrm{fit}/P_\mathrm{HFR}$)  for even configurations of neutral Nd. All parameters are in cm$^{-1}$.}
\lineup
%\begin{indented}
%\item[]
\begin{tabular}{@{}llrrlllrrllrr}
\\
\br
Param. $X$ & Cons. & $X_\mathrm{fit}$ & $f_X$ & & Param. $X$ & Cons. & $X_\mathrm{fit}$ & $f_X$ & & Cons. & $X_\mathrm{fit}$ & $f_X$ \\ \cline{1-13} \\
 & \multicolumn{3}{c}{A-6s2} & & & \multicolumn{3}{c}{A-ds} & &
\multicolumn{3}{c}{B-d6sp} \\
\cline{2-4} \cline{7-9} \cline{11-13}\\

E$_\mathrm{av}$ &  & 29612 & & & E$_\mathrm{av}$ &  & 43472 & & &  & 61355 &  \\ 
F$^2$(4f 4f) & r$_1$ & 67945 & 0.740 & & F$^2$(4f 4f) & fix & 68255 & 0.750 & & fix & 86247 & 0.853 \\
F$^4$(4f 4f) & r$_2$ & 38310 & 0.670 & & F$^4$(4f 4f) & fix & 42450 & 0.750 & & fix & 37856 & 0.597 \\
F$^6$(4f 4f) & r$_3$ & 28534 & 0.696 & & F$^6$(4f 4f) & fix & 30437 & 0.750 & & fix & 35027 & 0.769 \\
$\alpha$ & fix & 37 & & & $\alpha$ & fix & 37 & & & r$_{51}$ & 97 &  \\
$\beta$ & fix & \-963 & & & $\beta$ & fix & \-963 & & & fix & \-655 & \\
$\gamma$ & fix & 478 & & & $\gamma$ & fix & 478 & & & fix & 1691 & \\
$\zeta_{4f}$ & r$_4$ & 770 & 0.912 & & $\zeta_{4f}$ & r$_4$ & 765 & 0.912 & & r$_4$ & 975 & 1.032 \\
& & & & & $\zeta_{5d}$ & r$_4$ & 353 & 0.912 & & r$_4$ & 736 & 1.032\\
& & & & & $\zeta_{6p}$ & & & & & r$_4$ & 868 & 1.032\\
& & & & & F$^1$(4f 5d) & & & & & r$_9$ & 1854 &  \\
& & & & & F$^2$(4f 5d) & r$_1$ & 12316 & 0.740 & & r$_1$ & 27733 & 1.171 \\
& & & & & F$^3$(4f 5d) & & & & & r$_9$ & 1854 & \\
& & & & & F$^4$(4f 5d) & r$_2$ & 5307 & 0.670 & & r$_2$ & 31253 & 2.71 \\
& & & & & F$^1$(4f 6p) & & & & & r$_5$ & 613 & \\
& & & & & F$^2$(4f 6p) & & & & & r$_1$ & 4730 & 1.171 \\
& & & & & F$^1$(5d 6p) & & & & & r$_5$ & 613 & \\
& & & & & F$^2$(5d 6p) & & & & & r$_5$ & 16009 & 1.171 \\
& & & & & G$^1$(4f 5d) & r$_5$ & 5393 & 0.584 & & r$_6$ & 13100 & 1.147 \\
& & & & & G$^2$(4f 5d) & r$_9$ & 207 & & & & & \\
& & & & & G$^3$(4f 5d) & r$_5$ & 3868 & 0.584 & & r$_6$ & 10316 & 1.147 \\
& & & & & G$^4$(4f 5d) & r$_9$ & 1562 & & & & & \\
& & & & & G$^5$(4f 5d) & r$_5$ & 2832 & 0.584 & & r$_6$ & 7794 & 1.147 \\
& & & & & G$^3$(4f 6s) & r$_5$ & 947 & 0.584 & & r$_6$ & 2111 & 1.147 \\
& & & & & G$^2$(4f 6p) & & & & & r$_7$ & 1073 & 1.175 \\
& & & & & G$^4$(4f 6p) & & & & & r$_7$ & 682 & 0.842 \\
& & & & & G$^2$(5d 6s) & r$_5$ & 9719 & 0.584 & & r$_7$ & 17957 & 1.176 \\
& & & & & G$^1$(5d 6p) & & & & & r$_6$ & 9118 & 1.147 \\
& & & & & G$^3$(5d 6p) & & & & & r$_6$ & 6613 & 1.147 \\
& & & & & G$^1$(6s 6p) & & & & & r$_6$ & 26970 & 1.147 \\
\br
\end{tabular}
%\end{indented}
\end{table*}

When identifying the levels and trying to find the corresponding counter-experimental levels to the theoretical ones calculated by us for the least-square fit, we noticed some differences in the Land\'e g-factors for some levels. Figure \ref{fig:Lande_g_diff} shows that for most levels, the difference in Land\'e g-factors is limited to the region~[-0.1:0.1]. However, there are levels for which the absolute value of the difference exceeds 0.4. 
There are three such levels: the Land\'e g-factor of level $J = 3$ of configuration A-6s2, with an energy value of 11129~cm$^{-1}$, differs from its counter-experimental level by 1.237.
The $J = 5$ level of the A-ds configuration with an energy value of 21899~cm$^{-1}$ has a Land\'e g-factor that differs from the experimental one by 0.790. And, finally, the Land\'e g-factor of the level $J = 6$ of the A-ds configuration with an energy value of 11134~cm$^{-1}$ diverges from the experimental one by -0.451.

When the optimal set of parameters and the best (smallest) standard deviation are found, it is interesting to calculate the scaling factors (SF) for all parameters and groups of parameters that participated in the calculations, including CI ones.
Table \ref{tab:Nd_SF_even} shows the optimal parameters ($X_\mathrm{fit}$) for even parity configurations, as well as their constraints and scaling factors ($f_X$) if the parameter had an initial HFR value. Table \ref{tab:Nd_SF_odd} presents the same information for odd parity configurations, and table \ref{tab:Nd_SF_CI} for the CI parameters of even and odd parity configuration pairs.

Table \ref{tab:Nd_SF_even}--\ref{tab:Nd_SF_CI} also presents the constraints defining groups of fitting parameters: the parameters having the same $r_n$ value belong to the same group. Because our fit was made in several steps, in which the constraints have not been the same, the parameters with the same $r_n$ coefficients do not necessarily have the same scaling factors. Among the latter, we note especially large values for $G^k$ parameters of the 4f$^4$\,6s\,6p configuration and small values for CI parameters for even configuration pairs implying 4f$^3$\,5d\,6s\,6p. We can compare our fitted parameters to Ref.~\cite{aufmuth1992} which is dedicated to even-parity configurations 4f$^4$\,6s$^2$ + 4f$^4$\,5d\,6s. The agreement between theoretical and experimental levels is very good, but we note surprisingly small values of $F^k$(4f\,4f) parameters of the 4f$^4$\,5d\,6s configuration.

%\begin{table}
%\centering
%\caption{ \label{tab:Nd_odd_SF1} Parameters, constraints, fitted parameters and their scaling factors (F$_s$ = $P_{fit}$ / $P_{HFR}$) for A-6sp odd configuration of neutral Nd. All parameters are in cm$^{-1}$.}
%\lineup
%\begin{indented}
%\begin{tabular}{@{}llll}
%\\
%\br
%Param. P & Cons. & P$_{fit}$ & F$_s$ \\ \cline{1-4} \\

%E$_{av}$ &  & 52188 &  \\ 
%F$^2$(4f 4f) & r$_1$ & 72210 & 0.785 \\
%F$^4$(4f 4f) & r$_1$ & 45150 & 0.789 \\
%F$^6$(4f 4f) & r$_1$ & 32379 & 0.789 \\
%alpha & r$_{58}$ & 237 &  \\
%beta & r$_{58}$ & \-159 & \\
%gamma & r$_{58}$ & 411 & \\
%$\zeta_{4f}$ & r$_4$ & 828 & 0.980 \\
%$\zeta_{6p}$ & r$_4$ & 699 & 0.980 \\
%F$^1$(4f 6p) & r$_3$ & 1742 & \\
%F$^2$(4f 6p) & r$_3$ & 2355 & 0.593 \\
%G$^3$(4f 6s) & r$_6$ & 6463 & 3.384 \\
%G$^2$(4f 6p) & r$_7$ & 4754 & 5.185 \\
%G$^3$(4f 6p) & r$_7$ & 3842 & \\
%G$^4$(4f 6p) & r$_7$ & 2702 & 3.357 \\
%G$^1$(6s 6p) & r$_7$ & 17539 & 0.783 \\
%\br
%\end{tabular}
%%\end{indented}
%\end{table}

\begin{table*}
\centering
\caption{ \label{tab:Nd_SF_odd} Parameter names, constraints, fitted values and scaling factors ($f_X = X_\mathrm{fit} / P_\mathrm{HFR}$) for odd configurations of neutral Nd. All parameters are in cm$^{-1}$.}
\lineup
%\begin{indented}
%\item[]
\begin{tabular}{@{}llrrlllrrllrr}
\\
\br
Param. $X$ & Cons. & $X_\mathrm{fit}$ & $f_X$ & & Param. $X$ & Cons. & $X_\mathrm{fit}$ & $f_X$ & & Cons. & $X_\mathrm{fit}$ & $f_X$ \\ \cline{1-13} \\
& \multicolumn{3}{c}{A-6sp} & & & \multicolumn{3}{c}{B-ds2} & &
\multicolumn{3}{c}{B-d2s} \\
\cline{2-4} \cline{7-9} \cline{11-13}\\

E$_\mathrm{av}$ &  & 52188 & & & E$_\mathrm{av}$ &  & 32792 &  & & & 40755 &  \\ 
F$^2$(4f 4f) & r$_1$ & 72210 & 0.785 & & F$^2$(4f 4f) & r$_1$ & 70314 & 0.696 & & r$_1$ & 69931 & 0.696 \\
F$^4$(4f 4f) & r$_1$ & 45150 & 0.789 & & F$^4$(4f 4f) & r$_1$ & 36982 & 0.584 & & r$_1$ & 36765 & 0.584 \\
F$^6$(4f 4f) & r$_1$ & 32379 & 0.789 & & F$^6$(4f 4f) & r$_1$ & 21619 & 0.475 & & r$_1$ & 21489 & 0.475 \\
$\alpha$ & r$_{58}$ & 237 & & & $\alpha$ & r$_8$ & 73 & & & r$_8$ & 73&  \\
$\beta$ & r$_{58}$ & \-159 & & & $\beta$ & r$_8$ & \-667 & & & r$_8$ & \-667 & \\
$\gamma$ & r$_{58}$ & 411 & & & $\gamma$ & r$_8$ & 1744 & & & r$_8$ & 1744 & \\
& & & & & F$^2$(5d 5d) &  &  & & & r$_5$ & 19957 & 0.600 \\
& & & & & F$^4$(5d 5d) & & & & & r$_5$ & 10733 & 0.501\\
& & & & & $\alpha$ & & & & & r$_8$ & 71 &  \\
& & & & & $\beta$ & & & & & r$_8$ & \-650 & \\
$\zeta_{4f}$ & r$_4$ & 828 & 0.980 & & $\zeta_{4f}$ & r$_4$ & 881 & 0.932 & & r$_4$ & 877 & 0.932 \\
$\zeta_{6p}$ & r$_4$ & 699 & 0.980 & & $\zeta_{5d}$ & r$_4$ & 523 & 0.767 & & r$_4$ & 443 & 0.767\\
F$^1$(4f 6p) & r$_3$ & 1742 & & & F$^2$(4f 5d) & r$_2$ & 13678 & 0.598 & & r$_2$ & 12185 & 0.598 \\
F$^2$(4f 6p) & r$_3$ & 2355 & 0.593 & & F$^4$(4f 5d) & r$_2$ & 5523 & 0.499 & & r$_2$ & 4846 & 0.499 \\
G$^3$(4f 6s) & r$_6$ & 6463 & 3.384 & & G$^1$(4f 5d) & r$_6$ & 6267 & 0.570 & & r$_6$ & 5583 & 0.570 \\
G$^2$(4f 6p) & r$_7$ & 4754 & 5.185 & & G$^3$(4f 5d) & r$_6$ & 4921 & 0.570 & & r$_6$ & 4323 & 0.570 \\
G$^3$(4f 6p) & r$_7$ & 3842 & & & G$^5$(4f 5d) & r$_6$ & 3714 & 0.570 & & r$_6$ & 3248 & 0.570 \\
G$^4$(4f 6p) & r$_7$ & 2702 & 3.357 & & G$^3$(4f 6s) & & & & & r$_7$ & 866 & 0.566 \\
G$^1$(6s 6p) & r$_7$ & 17539 & 0.783 & & G$^2$(5d 6s) & & & & & r$_7$ & 8719 & 0.566 \\ 
\br
\end{tabular}
%\end{indented}
\end{table*}

%\newpage
\begin{table*}
\centering
\caption{\label{tab:Nd_SF_CI} Fitted configuration interaction (CI) parameters, their scaling factors ($f_X$ = $X_\mathrm{fit}$ / $X_\mathrm{HFR}$) and constraints for even and odd configurations of neutral Nd. All parameters are in cm$^{-1}$.}
\lineup
%\begin{indented}
%\item[]
\begin{tabular}{lrrllrr}
\\
\br
Parameter $X$ & $X_\mathrm{fit}$ & $f_X$ & & Parameter $X$ & $X_\mathrm{fit}$ & $f_X$ \\ \cline{1-3} \cline{4-7}\\
 & \multicolumn{2}{c}{\textbf{A-6s2 \textendash A-ds}} & & & \multicolumn{2}{c}{\textbf{A-6sp \textendash B-ds2}} \\
\cline{2-3} \cline{6-7} \\

R$^2$ (4f 6s, 4f 5d) & \-1074 & 0.441 & & R$^1$ (4f 6p, 5d 6s) & \-4065 & 0.475 \\
R$^3$ (4f 6s, 4f 5d) & 231 & 0.441 & & R$^3$ (4f 6p, 5d 6s) & \-866 & 0.475 \\
\\
 & \multicolumn{2}{c}{\textbf{A-6s2 \textendash B-d6sp}} & & & \multicolumn{2}{c}{\textbf{A-6sp \textendash B-d2s}} \\
 \cline{2-3} \cline{6-7} \\
R$^1$ (4f 6s, 5d 6p) & \-1517 & 0.163 & & R$^1$ (4f 6p, 5d 5d) & 1464 & 0.347 \\
R$^3$ (4f 6s, 5d 6p) & \-260 & 0.163 & & R$^3$ (4f 6p, 5d 5d) & 440 & 0.347 \\
\\
 & \multicolumn{2}{c}{\textbf{A-ds \textendash B-d6sp}} & & & \multicolumn{2}{c}{\textbf{B-ds2 \textendash B-d2s}} \\
 \cline{2-3} \cline{6-7} \\
R$^2$ (4f 4f, 4f 6p) & \-531 & 0.163 & & R$^2$ (4f 6s, 4f 5d) & \-628 & 0.487 \\
R$^4$ (4f 4f, 4f 6p) & \-348 & 0.163 & & R$^3$ (4f 6s, 4f 5d) & 607 & 0.487 \\
R$^1$ (4f 5d, 5d 6p) & 1047 & 0.163 & & R$^2$ (5d 6s, 5d 5d) & \-9305 & 0.487 \\
R$^3$ (4f 5d, 5d 6p) & 354 & 0.163 & & & & \\
R$^2$ (4f 5d, 5d 6p) & 27 & 0.163 & & & & \\
R$^4$ (4f 5d, 5d 6p) & 58 & 0.164 & & & & \\
\br
\end{tabular}
%\end{indented}
\end{table*}

%\begin{table}
%\centering
%\caption{\label{tab:Nd_odd_CI} Fitted configuration interaction (CI) parameters, their scaling factors (F$_s$ = $P_{fit}$ / $P_{HFR}$) and constraints for odd configurations of neutral Nd. All parameters are in cm$^{-1}$.}
%\lineup
%\begin{indented}
%\item[]
%\begin{tabular}{ccc}
%\\
%\br
%Parameter P & P$_{fit}$ & F$_s$  \\ \cline{1-3} \\
% & \multicolumn{2}{c}{\textbf{A-6sp \textendash B-ds2}} \\
% \cline{2-3}\\
%R$^1$ (4f 6p, 5d 6s) & \-4065 & 0.475 \\
%R$^3$ (4f 6p, 5d 6s) & \-866 & 0.475 \\
%\\
% & \multicolumn{2}{c}{\textbf{A-6sp \textendash B-d2s}} \\
% \cline{2-3}\\
%R$^1$ (4f 6p, 5d 5d) & 1464 & 0.347 \\
%R$^3$ (4f 6p, 5d 5d) & 440 & 0.347 \\
%\\
% & \multicolumn{2}{c}{\textbf{B-ds2 \textendash B-d2s}} \\
% \cline{2-3}\\
%R$^2$ (4f 6s, 4f 5d) & \-628 & 0.487 \\
%R$^3$ (4f 6s, 4f 5d) & 607 & 0.487 \\
%R$^2$ (5d 6s, 5d 5d) & \-9305 & 0.487 \\
%\br
%\end{tabular}
%\end{indented}
%\end{table}

\section{Conclusion}
\label{sec:conclusion}

In this article, we have given a theoretical interpretation of the spectrum of neutral neodymium, which is an essential component for new experiments with ultracold dipolar gases.
We did the calculations for three even configurations: 4f$^4$\,6s$^2$, 4f$^4$\,5d\,6s, 4f$^3$\,5d\,6s\,6p, and three odd configurations: 4f$^4$\,6s\,6p, 4f$^3$\,5d\,6s$^2$ and 4f$^3$\,5d$^2$\,6s. For this purpose we used Cowan's suite of codes.

Although Nd is a difficult element for such calculations, due to its very dense spectrum, we have been able to carry out the calculations by introducing a method in which we divide the calculation of each parity into two parts. The challenging part of this calculation was the least squares fit, because we needed to find experimental analogs for each theoretical level to which they should converge.
We were able to interpret more than 200 levels for odd parity configurations and 25 levels for even parity configurations, for which there were no detailed information in the NIST ASD database.
In the course of calculations, we noticed discrepancies with the NIST database values, for example, in Land\'e g-factors. After comparison we showed that for all levels except for three, the absolute value of the difference between the theoretical and experimental Land\'e g values does not exceed 0.4.

The logical continuation and perspective of this work for the future will be the calculation of the transition dipole moments (TDMs) and Einstein coefficients, which are necessary to characterize the efficiency of laser cooling and trapping of atoms.
For better accuracy, we plan to fit the Einstein coefficients using the FitAik package \cite{lepers2022}, for which we will use the optimal set of parameters that we have determined in this study.

\section*{Acknowledgements}
\label{sec:acknowledgements}

We acknowledge support from the NeoDip project (ANR-19-CE30-0018-01 from ``Agence Nationale de la Recherche''). M.L. also acknowledges the financial support of {}``R{\'e}gion Bourgogne Franche Comt{\'e}'' under the projet 2018Y.07063 {}``Th{\'e}CUP''.
Calculations have been performed using HPC resources from DNUM CCUB (Centre de Calcul de l'Universit\'e de Bourgogne).

\newpage
\section*{References}

\bibliography{Nd_article}

\providecommand{\newblock}{}
\begin{thebibliography}{10}
\expandafter\ifx\csname url\endcsname\relax
  \def\url#1{{\tt #1}}\fi
\expandafter\ifx\csname urlprefix\endcsname\relax\def\urlprefix{URL }\fi
\providecommand{\eprint}[2][]{\url{#2}}

\bibitem{baranov2008}
Baranov M~A 2008 {\em Physics Reports\/} {\bf 464} 71--111

\bibitem{bloch2008}
Bloch I, Dalibard J and Zwerger W 2008 {\em Reviews of modern physics\/} {\bf
  80} 885

\bibitem{bloch2012}
Bloch I, Dalibard J and Nascimbene S 2012 {\em Nature Physics\/} {\bf 8}
  267--276

\bibitem{griesmaier2005}
Griesmaier A, Werner J, Hensler S, Stuhler J and Pfau T 2005 {\em Physical
  Review Letters\/} {\bf 94} 160401

\bibitem{beaufils2008}
Beaufils Q, Chicireanu R, Zanon T, Laburthe-Tolra B, Mar{\'e}chal E, Vernac L,
  Keller J~C and Gorceix O 2008 {\em Physical Review A\/} {\bf 77} 061601

\bibitem{lepers2018}
Lepers M, Li H, Wyart J~F, Qu{\'e}m{\'e}ner G and Dulieu O 2018 {\em Phys. Rev.
  Lett.\/} {\bf 121} 063201
  \urlprefix\url{https://doi.org/10.1103/PhysRevLett.121.063201}

\bibitem{norcia2021}
Norcia M~A and Ferlaino F 2021 {\em Nature Physics\/} {\bf 17} 1349--1357

\bibitem{chomaz2022}
Chomaz L, Ferrier-Barbut I, Ferlaino F, Laburthe-Tolra B, Lev B~L and Pfau T
  2022 {\em arXiv preprint arXiv:2201.02672\/}

\bibitem{frisch2012}
Frisch A, Aikawa K, Mark M, Rietzler A, Schindler J, Zupani{\v{c}} E, Grimm R
  and Ferlaino F 2012 {\em Physical Review A\/} {\bf 85} 051401

\bibitem{seo2020efficient}
Seo B, Chen P, Chen Z, Yuan W, Huang M, Du S and Jo G~B 2020 {\em Physical
  Review A\/} {\bf 102} 013319

\bibitem{ban2005laser}
Ban H, Jacka M, Hanssen J~L, Reader J and McClelland J~J 2005 {\em Optics
  Express\/} {\bf 13} 3185--3195

\bibitem{berglund2008narrow}
Berglund A~J, Hanssen J~L and McClelland J~J 2008 {\em Physical review
  letters\/} {\bf 100} 113002

\bibitem{lu2010trapping}
Lu M, Youn S~H and Lev B~L 2010 {\em Physical review letters\/} {\bf 104}
  063001

\bibitem{lunden2020enhancing}
Lunden W, Du L, Cantara M, Barral P, Jamison A~O and Ketterle W 2020 {\em
  Physical Review A\/} {\bf 101} 063403

\bibitem{lu2011spectroscopy}
Lu M, Youn S~H and Lev B~L 2011 {\em Physical Review A\/} {\bf 83} 012510

\bibitem{dreon2017optical}
Dreon D, Sidorenkov L~A, Bouazza C, Maineult W, Dalibard J and Nascimbene S
  2017 {\em Journal of Physics B: Atomic, Molecular and Optical Physics\/} {\bf
  50} 065005

\bibitem{miao2014magneto}
Miao J, Hostetter J, Stratis G and Saffman M 2014 {\em Physical Review A\/}
  {\bf 89} 041401

\bibitem{sukachev2010magneto}
Sukachev D, Sokolov A, Chebakov K, Akimov A, Kanorsky S, Kolachevsky N and
  Sorokin V 2010 {\em Physical Review A\/} {\bf 82} 011405

\bibitem{vishnyakova2014two}
Vishnyakova G, Kalganova E, Sukachev D, Fedorov S, Sokolov A, Akimov A,
  Kolachevsky N and Sorokin V 2014 {\em Laser Physics\/} {\bf 24} 074018

\bibitem{inoue2018magneto}
Inoue R, Miyazawa Y and Kozuma M 2018 {\em Physical Review A\/} {\bf 97} 061607

\bibitem{ilzhofer2018two}
Ilzh{\"o}fer P, Durastante G, Patscheider A, Trautmann A, Mark M and Ferlaino F
  2018 {\em Physical Review A\/} {\bf 97} 023633

\bibitem{cowan1981}
Cowan R~D 1981 {\em The {t}heory of {a}tomic {s}tructure and {s}pectra\/} 3
  (Univ of California Press)

\bibitem{kramida2019}
Kramida A 2019 {\em Atoms\/} {\bf 7} 64
  \urlprefix\url{https://www.mdpi.com/2218-2004/7/3/64}

\bibitem{NIST_ASD}
Kramida A, {Yu~Ralchenko}, Reader J and {and NIST ASD Team} 2019 {NIST Atomic
  Spectra Database (ver. 5.7.1), [Online]. Available:
  {\tt{https://physics.nist.gov/asd}} [2020, April 10]. National Institute of
  Standards and Technology, Gaithersburg, MD.}

\bibitem{mcguinness_cowan}
McGuinness C 2009
  \urlprefix\url{http://www.tcd.ie/Physics/people/Cormac.McGuinness/Cowan/}

\bibitem{martin1978}
Martin W~C, Zalubas R and Hagan L 1978 Atomic energy levels-the rare-earth
  elements. Tech. rep. NATIONAL STANDARD REFERENCE DATA SYSTEM

\bibitem{aufmuth1992}
Aufmuth P, Bernard A and Kopp E~G 1992 {\em Z. Phys. D\/} {\bf 23} 15--18
  \urlprefix\url{https://link.springer.com/article/10.1007/BF01436697}

\bibitem{lepers2022}
Lepers M, Dulieu O and Wyart J~F 2022 {\em arXiv preprint arXiv:2207.14001\/}

\end{thebibliography}
%\begin{Bibliography}{25}
%\end{Bibliography}
\end{document}

\author{Gohar\;Hovhannesyan}
\author{Maxence\;Lepers}
\affil{\small {Laboratoire Interdisciplinaire Carnot de Bourgogne, UMR 6303 CNRS Univ. Bourgogne Franche-Comt\'{e}, 9 Avenue Alain Savary, BP 47 870, F-21078 DIJON Cedex FRANCE}}
%\author{Gohar\;Hovhannesyan$^1$, Maxence\;Lepers$^1$}

%\address{$^1$ Laboratoire Interdisciplinaire Carnot de Bourgogne, UMR 6303 CNRS Univ. Bourgogne Franche-Comt\'{e}, 9 Avenue Alain Savary, BP 47 870, F-21078 DIJON Cedex FRANCE}
\maketitle

%\section{Introduction}

This supplementary material is designed to accompany an article in the Journal of Physics B: Atomic, Molecular and Optical Physics. In what follows, we will use the following abbreviations for even parity configurations: 4f$^4$\,6s$^2$ = A-6s2, 4f$^4$\,5d\,6s = A-ds and 4f$^3$\,5d\,6s\,6p = B-d6sp and for odd parity configurations: 4f$^4$\,6s\,6p = A-6sp, 4f$^3$\,5d\,6s$^2$ = B-ds2 and 4f$^3$\,5d$^2$\,6s = B-d2s.

\section{Fitted energy parameters}

In this section, we present tables with fitted parameter values ($X_\mathrm{fit}$) for configurations of Nd after the last calculation and their scaling factors ($f_X$) (if the parameter had an initial HFR value), when only NIST interpreted levels are included in the fitting process (see Subsection 3.2 of the main text).

%\newpage
\begin{center}
\begin{longtable}{@{\extracolsep{\fill}} llrllllrlllrl @{}}
%\centering
\caption*{\textbf{Table \ref{tab:Nd_SF_even}: } Parameters, constraints, fitted parameters and their scaling factors ($f_X$ = $X_\mathrm{fit}$ / $X_\mathrm{HFR}$) for even configurations of neutral Nd. All parameters are in cm$^{-1}$.}
\label{tab:Nd_SF_even} \\
\toprule
\endfirsthead
\caption* {\textbf{Table \ref{tab:Nd_SF_even} Continued} }\\
\toprule 
Param. $X$ & Cons. & $X_\mathrm{fit}$ & $f_X$ & & Param. $X$ & Cons. & $X_\mathrm{fit}$ & $f_X$ & & Cons. & X$_\mathrm{fit}$ & $f_X$ \\ \cmidrule{1-13} \\
 & \multicolumn{3}{c}{A-6s2} & & & \multicolumn{3}{c}{A-ds} & &
\multicolumn{3}{c}{B-d6sp} \\
\cmidrule{2-4} \cmidrule{7-9} \cmidrule{11-13}\\

\endhead
\cmidrule{1-13}
\endfoot
\endlastfoot
Param. $X$ & Cons. & $X_\mathrm{fit}$ & $f_X$ & & Param. $X$ & Cons. & $X_\mathrm{fit}$ & $f_X$ & & Cons. & $X_\mathrm{fit}$ & $f_X$ \\ \cmidrule{1-13} \\
 & \multicolumn{3}{c}{A-6s2} & & & \multicolumn{3}{c}{A-ds} & &
\multicolumn{3}{c}{B-d6sp} \\
\cmidrule{2-4} \cmidrule{7-9} \cmidrule{11-13}\\

E$_\mathrm{av}$ &  & 29054 & & & E$_\mathrm{av}$ &  & 43510 & & &  & 61333 &  \\ 
F$^2$(4f 4f) & r$_1$ & 67029 & 0.730 & & F$^2$(4f 4f) & fixed & 68255 & 0.750 & & fixed & 86247 & 0.853 \\
F$^4$(4f 4f) & r$_2$ & 36840 & 0.645 & & F$^4$(4f 4f) & fixed & 42450 & 0.750 & & fixed & 37856 & 0.597 \\
F$^6$(4f 4f) & r$_3$ & 25772 & 0.629 & & F$^6$(4f 4f) & fixed & 30437 & 0.750 & & fixed & 35027 & 0.769 \\
$\alpha$ & fixed & 37 & & & $\alpha$ & fixed & 37 & & & r$_{51}$ & 96 &  \\
$\beta$ & fixed & \-963 & & & $\beta$ & fixed & \-963 & & & fixed & \-655 & \\
$\gamma$ & fixed & 478 & & & $\gamma$ & fixed & 478 & & & fixed & 1691 & \\
$\zeta_{4f}$ & r$_4$ & 775 & 0.917 & & $\zeta_{4f}$ & r$_4$ & 770 & 0.918 & & r$_4$ & 981 & 1.039 \\
& & & & & $\zeta_{5d}$ & r$_4$ & 356 & 0.918 & & r$_4$ & 741 & 1.038\\
& & & & & $\zeta_{6p}$ & & & & & r$_4$ & 873 & 1.038\\
& & & & & F$^1$(4f 5d) & & & & & r$_9$ & 2111 &  \\
& & & & & F$^2$(4f 5d) & r$_1$ & 12150 & 0.730 & & r$_1$ & 27359 & 1.155 \\
& & & & & F$^3$(4f 5d) & & & & & r$_9$ & 2111 & \\
& & & & & F$^4$(4f 5d) & r$_2$ & 5104 & 0.645 & & r$_2$ & 30055 & 2.610 \\
& & & & & F$^1$(4f 6p) & & & & & r$_5$ & 619 & \\
& & & & & F$^2$(4f 6p) & & & & & r$_1$ & 4667 & 1.115 \\
& & & & & F$^1$(5d 6p) & & & & & r$_5$ & 619 & \\
& & & & & F$^2$(5d 6p) & & & & & r$_5$ & 15793 & 1.115 \\
& & & & & G$^1$(4f 5d) & r$_5$ & 5443 & 0.589 & & r$_6$ & 12950 & 1.134 \\
& & & & & G$^2$(4f 5d) & r$_9$ & 235 & & & & & \\
& & & & & G$^3$(4f 5d) & r$_5$ & 3904 & 0.586 & & r$_6$ & 10198 & 1.134 \\
& & & & & G$^4$(4f 5d) & r$_9$ & 1779 & & & & & \\
& & & & & G$^5$(4f 5d) & r$_5$ & 2858 & 0.589 & & r$_6$ & 7705 & 1.134 \\
& & & & & G$^3$(4f 6s) & r$_5$ & 956 & 0.589 & & r$_6$ & 2087 & 1.134 \\
& & & & & G$^2$(4f 6p) & & & & & r$_7$ & 1138 & 1.247 \\
& & & & & G$^4$(4f 6p) & & & & & r$_7$ & 723 & 0.893 \\
& & & & & G$^2$(5d 6s) & r$_5$ & 9809 & 0.589 & & r$_7$ & 19047 & 1.247 \\
& & & & & G$^1$(5d 6p) & & & & & r$_6$ & 6537 & 1.134 \\
& & & & & G$^3$(5d 6p) & & & & & r$_6$ & 6613 & 1.134 \\
& & & & & G$^1$(6s 6p) & & & & & r$_6$ & 26659 & 1.134 \\

\bottomrule
%\br
%\end{tabular}
%\end{indented}
\end{longtable}
\end{center}

\begin{center}
\begin{longtable}{@{\extracolsep{\fill}} llrllllrlllrl @{}}
%\centering
\caption*{\textbf{Table \ref{tab:Nd_SF_odd}: }Parameters, constraints, fitted parameters and their scaling factors ($f_X = X_\mathrm{fit}$ / $X_\mathrm{HFR}$) for odd configurations of neutral Nd. All parameters are in cm$^{-1}$.}
\label{tab:Nd_SF_odd} \\
\toprule
\endfirsthead
\caption* {\textbf{Table \ref{tab:Nd_SF_odd} Continued} }\\
\toprule
Param. $X$ & Cons. & $X_\mathrm{fit}$ & $f_X$ & & Param. $X$ & Cons. & $X_\mathrm{fit}$ & $f_X$ & & Cons. & $X_\mathrm{fit}$ & $f_X$ \\ \cmidrule{1-13} \\
& \multicolumn{3}{c}{A-6sp} & & & \multicolumn{3}{c}{B-ds2} & &
\multicolumn{3}{c}{B-d2s} \\
\cmidrule{2-4} \cmidrule{7-9} \cmidrule{11-13}\\

\endhead
\cmidrule{1-13}
\endfoot
\endlastfoot
Param. $X$ & Cons. & $X_\mathrm{fit}$ & $f_X$ & & Param. $X$ & Cons. & $X_\mathrm{fit}$ & $f_X$ & & Cons. & $X_\mathrm{fit}$ & $f_X$ \\ \cmidrule{1-13} \\
& \multicolumn{3}{c}{A-6sp} & & & \multicolumn{3}{c}{B-ds2} & &
\multicolumn{3}{c}{B-d2s} \\
\cmidrule{2-4} \cmidrule{7-9} \cmidrule{11-13}\\

E$_\mathrm{av}$ &  & 52415 & & & E$_\mathrm{av}$ &  & 32902 &  & & & 40943 &  \\ 
F$^2$(4f 4f) & r$_1$ & 72969 & 0.793 & & F$^2$(4f 4f) & r$_1$ & 71053 & 0.703 & & r$_1$ & 70665 & 0.703\\
F$^4$(4f 4f) & r$_1$ & 45624 & 0.797 & & F$^4$(4f 4f) & r$_1$ & 37370 & 0.590 & & r$_1$ & 37151 & 0.589 \\
F$^6$(4f 4f) & r$_1$ & 32719 & 0.797 & & F$^6$(4f 4f) & r$_1$ & 21846 & 0.479 & & r$_1$ & 21714 & 0.479 \\
$\alpha$ & r$_{58}$ & 255 & & & $\alpha$ & r$_8$ & 78 & & & r$_8$ & 78 &  \\
$\beta$ & r$_{58}$ & \-171 & & & $\beta$ & r$_8$ & \-717 & & & r$_8$ & \-717 & \\
$\gamma$ & r$_{58}$ & 441 & & & $\gamma$ & r$_8$ & 1875 & & & r$_8$ & 1875 & \\
& & & & & F$^2$(5d 5d) &  &  & & & r$_5$ & 20344 & 0.612 \\
& & & & & F$^4$(5d 5d) & & & & & r$_5$ & 10941 & 0.511\\
& & & & & $\alpha$ & & & & & r$_8$ & 76 &  \\
& & & & & $\beta$ & & & & & r$_8$ & \-698 & \\
$\zeta_{4f}$ & r$_4$ & 832 & 0.985 & & $\zeta_{4f}$ & r$_4$ & 885 & 0.937 & & r$_4$ & 881 & 0.937 \\
$\zeta_{6p}$ & r$_4$ & 702 & 0.985 & & $\zeta_{5d}$ & r$_4$ & 524 & 0.771 & & r$_4$ & 445 & 0.771\\
F$^1$(4f 6p) & r$_3$ & 1726 & & & F$^2$(4f 5d) & r$_2$ & 13821 & 0.604 & & r$_2$ & 12313 & 0.604 \\
F$^2$(4f 6p) & r$_3$ & 2333 & 0.588 & & F$^4$(4f 5d) & r$_2$ & 5580 & 0.505 & & r$_2$ & 4897 & 0.505 \\
G$^3$(4f 6s) & r$_6$ & 6406 & 3.354 & & G$^1$(4f 5d) & r$_6$ & 6212 & 0.565 & & r$_6$ & 5534 & 0.565 \\
G$^2$(4f 6p) & r$_7$ & 4762 & 5.194 & & G$^3$(4f 5d) & r$_6$ & 4878 & 0.565 & & r$_6$ & 4285 & 0.565 \\
G$^3$(4f 6p) & r$_7$ & 3848 & & & G$^5$(4f 5d) & r$_6$ & 3682 & 0.565 & & r$_6$ & 3219 & 0.565 \\
G$^4$(4f 6p) & r$_7$ & 2706 & 3.363 & & G$^3$(4f 6s) & & & & & r$_7$ & 867 & 0.567 \\
G$^1$(6s 6p) & r$_7$ & 17566 & 0.784 & & G$^2$(5d 6s) & & & & & r$_7$ & 8733 & 0.567 \\ 
%\br

\bottomrule

%\end{tabular}
%\end{indented}
\end{longtable}
\end{center}

\begin{center}
\begin{longtable}{@{\extracolsep{\fill}} lrrllrr @{}}
%\centering
\caption*{\textbf{Table \ref{tab:Nd_SF_CI}: }Fitted configuration interaction (CI) parameters, their scaling factors ($f_X$ = $X_\mathrm{fit}$ / $X_\mathrm{HFR}$) and constraints for even and odd configurations of neutral Nd. All parameters are in cm$^{-1}$.}
\label{tab:Nd_SF_CI}
%\lineup
%\begin{indented}
%\item[]
%\begin{tabular}{lllllll}
\\
\toprule
Parameter $X$ & $X_\mathrm{fit}$ & $f_X$ & & Parameter $X$ & $X_\mathrm{fit}$ & $f_X$ \\ \cline{1-3} \cline{5-7}\\
 & \multicolumn{2}{c}{\textbf{A-6s2 \textendash A-ds}} & & & \multicolumn{2}{c}{\textbf{A-6sp \textendash B-ds2}} \\
\cline{2-3} \cline{6-7} \\

R$^2$ (4f 6s, 4f 5d) & -1033 & 0.425 & & R$^1$ (4f 6p, 5d 6s) & -4049 & 0.473 \\
R$^3$ (4f 6s, 4f 5d) & 222 & 0.425 & & R$^3$ (4f 6p, 5d 6s) & -863 & 0.474 \\
\\
 & \multicolumn{2}{c}{\textbf{A-6s2 \textendash B-d6sp}} & & & \multicolumn{2}{c}{\textbf{A-6sp \textendash B-d2s}} \\
 \cline{2-3} \cline{6-7} \\
R$^1$ (4f 6s, 5d 6p) & -1496 & 0.161 & & R$^1$ (4f 6p, 5d 5d) & 1476 & 0.351 \\
R$^3$ (4f 6s, 5d 6p) & -257 & 0.161 & & R$^3$ (4f 6p, 5d 5d) & 443 & 0.351 \\
\\
 & \multicolumn{2}{c}{\textbf{A-ds \textendash B-d6sp}} & & & \multicolumn{2}{c}{\textbf{B-ds2 \textendash B-d2s}} \\
 \cline{2-3} \cline{6-7} \\
R$^2$ (4f 4f, 4f 6p) & -524 & 0.161 & & R$^2$ (4f 6s, 4f 5d) & -595 & 0.461 \\
R$^4$ (4f 4f, 4f 6p) & -343 & 0.161 & & R$^3$ (4f 6s, 4f 5d) & 575 & 0.461 \\
R$^1$ (4f 5d, 5d 6p) & 1033 & 0.161 & & R$^2$ (5d 6s, 5d 5d) & -8813 & 0.461 \\
R$^3$ (4f 5d, 5d 6p) & 349 & 0.161 & & & & \\
R$^2$ (4f 5d, 5d 6p) & 26 & 0.161 & & & & \\
R$^4$ (4f 5d, 5d 6p) & 57 & 0.161 & & & & \\
\bottomrule
%\end{tabular}
%\end{indented}
\end{longtable}
\end{center}

\newpage
\section{Identified levels}

Here we present detailed tables for newly interpreted levels for even and odd parity configurations of neutral Nd. In what follows we use the abbreviations: conf.=configuration, exp.=experimental data taken from the NIST database and dom. term=dominant eigenvalue term with its percentage.

\begin{center}
\begin{longtable}{ l l l l l l r r r r @{}}

\caption*{\textbf{Table \ref{Nd_even}: } Newly interpreted levels and their comparison with NIST database values for even parity. All energy levels are in cm$^{-1}$.}
\label{Nd_even} \\
\toprule
\endfirsthead
\caption* {\textbf{Table \ref{Nd_even} Continued} }\\
\toprule

Conf. & $J$ & \multicolumn{2}{c}{Energy } & & \multicolumn{2}{c}{Land\'e g } & & \multicolumn{2}{c}{Dom. term} \\ \cmidrule{1-10} \\
& & Theory & Exp. & & Theory & Exp. & & \\
 \cmidrule{3-4} \cmidrule{6-7} \\

\endhead
\cmidrule{1-9}
\endfoot
%\bottomrule
\endlastfoot
Conf. & $J$ & \multicolumn{2}{c}{Energy } & & \multicolumn{2}{c}{Land\'e g } & & \multicolumn{2}{c}{Dom. term} \\ \cmidrule{1-10} \\
& & Theory & Exp. & & Theory & Exp. & & \\
 \cmidrule{3-4} \cmidrule{6-7} \\
B-d6sp & 4 &  23329 & 23347 & & 0.744 & 0.740 & & $^7$I & 62\% \\
A-ds   & 4 &  23840 & 23808 & & 1.391 & 0.610 & & $^5$D & 39\% \\
A-ds   & 5 &  24555 & 24578 & & 0.926 & 0.870 & & $^7$I & 22\% \\
A-6s2  & 6 &  23113 & 23109 & & 1.028 & 0.895 & & $^3$I & 69\% \\
A-ds   & 6 &  23093 & 23222 & & 1.002 & 0.965 & & $^5$K & 14\% \\
A-ds   & 6 &  23331 & 23242 & & 1.373 & 1.115 & & $^7$F & 60\% \\
A-ds   & 6 &  24950 & 24927 & & 1.062 & 0.955 & & $^5$K & 15\% \\
B-d6sp & 6 &  25166 & 25207 & & 1.023 & 0.905 & & $^7$I & 42\% \\
B-d6sp & 6 &  25733 & 25746 & & 0.905 & 0.860 & & $^5$L & 10\% \\
A-ds   & 7 &  24396 & 24260 & & 1.063 & 1.010 & & $^5$L & 14\% \\
A-ds   & 7 &  24773 & 24821 & & 1.233 & 1.060 & & $^5$H & 36\% \\
B-d6sp & 7 &  25037 & 25005 & & 0.906 & 0.905 & & $^3$L & 30\% \\
A-ds   & 7 &  25406 & 25384 & & 1.086 & 0.971 & & $^5$K & 18\% \\
B-d6sp & 7 &  26444 & 26516 & & 1.043 & 1.085 & & $^7$I & 12\% \\
A-ds   & 7 &  26871 & 26898 & & 1.184 & 1.040 & & $^5$I & 16\% \\
A-ds   & 7 &  27038 & 27000 & & 1.148 & 1.055 & & $^7$I & 17\% \\
B-d6sp & 8 &  23919 & 23953 & & 1.080 & 1.053 & & $^7$L & 64\% \\
A-ds   & 8 &  25177 & 25190 & & 1.096 & 1.110 & & $^5$K & 25\% \\
B-d6sp & 8 &  25572 & 25529 & & 1.184 & 0.990 & & $^7$K & 62\% \\
A-ds   & 8 &  26472 & 26425 & & 1.013 & 1.135 & & $^1$L & 22\% \\
A-ds   & 8 &  27451 & 27324 & & 1.109 & 1.040 & & $^5$I & 25\% \\
A-ds   & 8 &  27660 & 27815 & & 0.951 & 1.015 & & $^5$N & 18\% \\
B-d6sp & 8 &  27997 & 27922 & & 1.158 & 1.125 & & $^7$K & 38\% \\
A-ds   & 9 &  26928 & 26987 & & 1.077 & 1.140 & & $^1$M & 24\% \\
A-ds   & 9 &  27418 & 27266 & & 1.036 & 1.040 & & $^1$M & 23\% \\

\bottomrule
\end{longtable}
\end{center}

\begin{center}
\begin{longtable}{ l l l l l l l r r @{}}

\caption*{\textbf{Table \ref{Nd_J_3}: }Identification and comparison with NIST database values for $J = 3$ levels of odd parity configurations. All energy levels are in cm$^{-1}$.}
\label{Nd_J_3} \\
\toprule
\endfirsthead
%\multicolumn{4}{@{}l}{\ldots continued}\\\hline

\caption* {\textbf{Table \ref{Nd_J_3} Continued} }\\\toprule
Conf. & \multicolumn{2}{c}{Energy } & & \multicolumn{2}{c}{Land\'e g } & & \multicolumn{2}{c}{Dom. term} \\ \cmidrule{1-9} \\
& Theory & Exp. & & Theory & Exp. & & \\
\cmidrule{2-3} \cmidrule{5-6} \\
\endhead
%\multicolumn{4}{@{}l}{\ldots continued}\\\hline

\cmidrule{1-9}
\endfoot
%\bottomrule
\endlastfoot
Conf. & \multicolumn{2}{c}{Energy } & & \multicolumn{2}{c}{Land\'e g } & & \multicolumn{2}{c}{Dom. term} \\ \cmidrule{1-9} \\

 & Theory & Exp. & & Theory & Exp. & & \\
\cmidrule{2-3} \cmidrule{5-6} \\
%\endhead
B-d2s &  16924 & 16974 & & 0.685 & 0.520 & & $^7$H & 49\% \\
B-d2s &  20303 & 20282 & & 0.302 & 0.895 & & $^7$I & 77\% \\
B-ds2 &  21139 & 21228 & & 1.259 & 1.032 & & $^5$D & 26\% \\
B-d2s &  21890 & 21952 & & 0.974 & 1.070 & & $^7$G & 33\% \\
B-d2s &  22208 & 22229 & & 0.817 & 0.705 & & $^5$H & 36\% \\
B-d2s &  22504 & 22491 & & 0.766 & 0.830 & & $^5$H & 24\% \\
B-d2s &  22649 & 22631 & & 0.790 & 1.130 & & $^7$H & 18\% \\
B-d2s &  22888 & 22930 & & 1.012 & 0.765 & & $^3$G & 7\% \\
B-d2s &  23034 & 22956 & & 0.997 & 0.990 & & $^5$G & 14\% \\
B-ds2 &  23280 & 23218 & & 0.956 & 1.060 & & $^3$G & 15\% \\
B-ds2 &  23687 & 23762 & & 1.108 & 1.190 & & $^1$F & 15\% \\
B-d2s &  24804 & 24775 & & 1.180 & 1.352 & & $^5$G & 8\%  \\
B-d2s &  25061 & 25164 & & 1.088 & 1.050 & & $^5$G & 16\% \\
A-6sp &  25331 & 25282 & & 1.209 & 1.360 & & $^5$D & 10\% \\
A-6sp &  25535 & 25500 & & 0.904 & 0.825 & & $^7$H & 44\% \\
B-d2s &  25669 & 25642 & & 1.117 & 0.900 & & $^3$F & 8\% \\
B-d2s &  25754 & 25788 & & 1.166 & 0.900 & & $^7$G & 26\% \\
B-d2s &  25996 & 26061 & & 1.134 & 1.360 & & $^7$G & 12\% \\
B-d2s &  26170 & 26096 & & 0.759 & 0.980 & & $^7$I & 24\% \\
B-d2s &  26219 & 26163 & & 1.057 & 1.180 & & $^7$S & 4\% \\
B-ds2 &  26283 & 26346 & & 1.065 & 1.025 & & $^3$D & 4\% \\
B-d2s &  26359 & 26395 & & 0.880 & 1.060 & & $^7$I & 13\% \\
A-6sp &  26425 & 26463 & & 1.133 & 0.865 & & $^7$G & 49\% \\

%***
%B-d2s & 20306.0 & 20281.9 & 0.304 & 0.895 & 7I 77\% \\
%B-d2s & 21892.4 & 21951.8 & 0.970 & 1.070 & 7G 31\% \\
%B-d2s & 22495.8 & 22490.9 & 0.726 & 0.830 & 5H 26\% \\
%B-d2s & 23021.9 & 22956.5 & 0.953 & 0.990 & 5G 15\% \\
%B-ds2 & 23260.6 & 23217.9 & 0.962 & 1.060 & 3G 15\% \\
%B-d2s & 23680.5 & 23761.8 & 1.118 & 1.190 & 1F 22\% \\
%B-d2s & 24802.6 & 24775.1 & 1.211 & 1.352 & 5D 8\% \\
%B-d2s & 25062.3 & 25164.4 & 1.067 & 1.050 & 5G 26\% \\
%A-6sp & 25535.8 & 25499.9 & 0.891 & 0.825 & 7H 47\% \\
%B-d2s & 26222.3 & 26162.8 & 1.055 & 1.180 & 7S 4\% \\
%B-d2s & 26281.1 & 26345.5 & 1.073 & 1.025 & 3D 8\% \\
%A-6sp & 26415.8 & 26395.0 & 1.098 & 1.060 & 7G 47\% \\

\bottomrule
\end{longtable}
\end{center}

%\newpage

%\captionsetup{type=table}

\begin{center}
\begin{longtable}{ l l l l l l l l r @{}}
%\begin{longtable}{\linewidth}{ @{\extracolsep{\fill}} ll *{7}l @{}}

\caption*{\textbf{Table \ref{Nd_J_4}: }Identification and comparison with NIST database values for $J = 4$ levels of odd parity configurations. All energy levels are in cm$^{-1}$.}
\label{Nd_J_4} \\
\toprule
\endfirsthead
\caption* {\textbf{Table \ref{Nd_J_4} Continued} }\\
\toprule

Conf. & \multicolumn{2}{c}{Energy } & & \multicolumn{2}{c}{Land\'e g } & & \multicolumn{2}{c}{Dom. term} \\ \cmidrule{1-9} \\
& Theory & Exp. & & Theory & Exp. & & \\
\cmidrule{2-3} \cmidrule{5-6} \\

\endhead
\cmidrule{1-9}
\endfoot
%\bottomrule
\endlastfoot
Conf. & \multicolumn{2}{c}{Energy } & & \multicolumn{2}{c}{Land\'e g } & & \multicolumn{2}{c}{Dom. term} \\ \cmidrule{1-9} \\
 & Theory & Exp. & & Theory & Exp. & &\\
 \cmidrule{2-3} \cmidrule{5-6} \\
B-d2s &  17073 & 17032 & & 1.231 & 1.020 & & $^7$F & 26\% \\
B-d2s &  17206 & 17320 & & 0.983 & 0.865 & & $^7$H & 33\% \\
B-ds2 &  18518 & 18436 & & 0.954 & 1.075 & & $^3$H & 21\% \\
B-ds2 &  19232 & 19209 & & 0.888 & 0.990 & & $^5$G & 12\% \\
A-6sp &  19569 & 19590 & & 1.124 & 0.785 & & $^7$G & 33\% \\
B-d2s &  19770 & 19770 & & 0.873 & 0.920 & & $^5$I & 15\% \\
A-6sp &  19952 & 19957 & & 0.798 & 0.910 & & $^5$I & 23\% \\
B-ds2 &  19984 & 20047 & & 0.917 & 0.900 & & $^3$G & 12\% \\
B-d2s &  20960 & 20860 & & 0.886 & 1.080 & & $^7$I & 35\% \\
A-6sp &  20995 & 21009 & & 1.611 & 1.280 & & $^7$D & 89\% \\
B-d2s &  21125 & 21185 & & 0.827 & 0.920 & & $^5$I & 10\% \\
B-d2s &  21297 & 21314 & & 0.916 & 0.985 & & $^5$G & 16\% \\
B-d2s &  21448 & 21488 & & 0.909 & 0.910 & & $^5$I & 12\% \\
B-ds2 &  22038 & 22077 & & 1.046 & 1.035 & & $^3$F & 10\% \\
B-ds2 &  22392 & 22471 & & 1.099 & 0.995 & & $^3$F & 12\% \\
B-d2s &  22584 & 22623 & & 1.003 & 1.135 & & $^3$H & 15\% \\
B-d2s &  22651 & 22678 & & 1.033 & 0.885 & & $^7$G & 9\% \\
B-ds2 &  22867 & 22815 & & 0.846 & 0.975 & & $^5$I & 23\% \\
B-ds2 &  22953 & 23016 & & 0.794 & 0.810 & & $^5$I & 28\% \\
B-d2s &  23026 & 23089 & & 0.941 & 1.250 & & $^7$H & 27\% \\
B-d2s &  23284 & 23204 & & 1.170 & 1.200 & & $^7$P & 11\% \\
B-d2s &  23444 & 23439 & & 0.998 & 0.885 & & $^7$H & 18\% \\
B-d2s &  23527 & 23563 & & 1.067 & 0.940 & & $^5$H & 7\% \\
B-d2s &  23877 & 23846 & & 0.989 & 1.175 & & $^5$I & 7\% \\
B-ds2 &  23962 & 23953 & & 0.899 & 0.830 & & $^3$G & 12\% \\
B-d2s &  24070 & 24001 & & 1.237 & 1.110 & & $^7$D & 16\% \\
B-ds2 &  24747 & 24674 & & 1.070 & 1.060 & & $^3$G & 11\% \\
B-d2s &  25352 & 25448 & & 1.148 & 1.175 & & $^7$D & 10\% \\
B-ds2 &  25494 & 25476 & & 0.881 & 0.915 & & $^5$H & 23\% \\
B-ds2 &  25621 & 25621 & & 1.085 & 1.175 & & $^5$H & 17\% \\
B-d2s &  25674 & 25641 & & 1.041 & 1.120 & & $^5$F & 7\% \\
B-ds2 &  25798 & 25791 & & 1.138 & 0.920 & & $^5$G & 18\% \\
B-d2s &  26182 & 26232 & & 1.083 & 1.025 & & $^5$F & 5\% \\
B-d2s &  26642 & 26662 & & 1.120 & 1.000 & & $^5$D & 5\% \\
B-d2s &  26734 & 26684 & & 1.172 & 1.130 & & $^5$D & 8\% \\
B-d2s &  26770 & 26770 & & 1.052 & 1.200 & & $^3$G & 7\% \\
B-ds2 &  26995 & 27144 & & 1.137 & 0.965 & & $^5$G & 14\% \\
B-d2s &  27261 & 27230 & & 0.972 & 1.125 & & $^7$I & 10\% \\
B-d2s &  27478 & 27490 & & 1.109 & 1.085 & & $^3$H & 10\% \\
A-6sp &  27678 & 27717 & & 1.038 & 0.890 & & $^3$G & 8\%  \\
B-d2s &  27801 & 27780 & & 1.059 & 1.030 & & $^3$G & 4\% \\
B-d2s &  27924 & 27860 & & 1.208 & 1.125 & & $^5$D & 9\% \\
B-d2s &  28004 & 27990 & & 1.166 & 0.930 & & $^5$F & 7\% \\

\bottomrule
\end{longtable}
\end{center}

%\newpage
%\captionsetup{type=table}

\begin{center}
\begin{longtable}{ l l l l l l l r r @{}}

\caption*{\textbf{Table \ref{Nd_J_5}: }Identification and comparison with NIST database values for $J = 5$ levels of odd parity configurations. All energy levels are in cm$^{-1}$.}
\label{Nd_J_5} \\
\toprule
\endfirsthead
\caption* {\textbf{Table \ref{Nd_J_5} Continued} }\\
\toprule

Conf. & \multicolumn{2}{c}{Energy } & & \multicolumn{2}{c}{Land\'e g } & & \multicolumn{2}{c}{Dom. term} \\ \cmidrule{1-9} \\
& Theory & Exp. & & Theory & Exp. & & \\
\cmidrule{2-3} \cmidrule{5-6} \\

\endhead
\cmidrule{1-9}
\endfoot
%\bottomrule
\endlastfoot
Conf. & \multicolumn{2}{c}{Energy } & & \multicolumn{2}{c}{Land\'e g } & & \multicolumn{2}{c}{Dom. term} \\ \cmidrule{1-9} \\
 & Theory & Exp. & & Theory & Exp. & & \\
 \cmidrule{2-3} \cmidrule{5-6} \\
 
A-6sp &  15963 & 16028 & & 1.000 & 0.915 & & $^7$I & 75\% \\
B-ds2 &  17826 & 17791 & & 0.907 & 1.005 & & $^3$I & 20\% \\
B-d2s &  17985 & 18030 & & 0.990 & 0.970 & & $^5$K & 13\% \\ 
B-d2s &  18060 & 18068 & & 1.142 & 0.920 & & $^7$H & 21\% \\
B-ds2 &  19180 & 19226 & & 0.944 & 0.944 & & $^5$H & 31\% \\
A-6sp &  19722 & 19648 & & 0.931 & 1.070 & & $^5$K & 12\% \\
A-6sp &  20186 & 20177 & & 1.229 & 0.960 & & $^7$G & 54\% \\
B-d2s &  20982 & 20963 & & 0.994 & 0.990 & & $^5$I & 20\% \\
B-ds2 &  21337 & 21272 & & 1.054 & 1.040 & & $^1$H & 9\% \\
B-d2s &  21746 & 21727 & & 1.063 & 1.000 & & $^5$F & 9\% \\
B-d2s &  22069 & 22129 & & 1.052 & 1.060 & & $^5$H & 7\% \\
B-ds2 &  22108 & 22192 & & 0.943 & 1.090 & & $^3$I & 12\% \\
B-d2s &  22415 & 22367 & & 1.211 & 1.085 & & $^7$G & 9\% \\
B-d2s &  22828 & 22737 & & 1.062 & 1.070 & & $^3$H & 13\% \\
B-ds2 &  22987 & 23050 & & 1.025 & 1.060 & & $^3$G & 11\% \\
B-d2s &  23262 & 23198 & & 1.053 & 1.085 & & $^7$H & 18\% \\
B-d2s &  23459 & 23434 & & 1.047 & 0.965 & & $^7$H & 15\% \\
B-d2s &  23598 & 23573 & & 1.083 & 1.080 & & $^5$H & 11\% \\
B-d2s &  23830 & 23830 & & 1.152 & 1.165 & & $^5$H & 6\% \\
B-ds2 &  24063 & 23968 & & 1.017 & 0.980 & & $^5$I & 12\% \\
B-d2s &  24266 & 24292 & & 1.037 & 1.060 & & $^5$H & 10\% \\
A-6sp &  24333 & 24364 & & 0.984 & 1.090 & & $^3$I & 19\% \\
B-d2s &  24443 & 24428 & & 0.981 & 1.190 & & $^3$I & 20\% \\
B-d2s &  24638 & 24586 & & 1.107 & 1.110 & & $^7$D & 8\% \\
B-d2s &  24743 & 24746 & & 1.245 & 1.185 & & $^7$F & 17\% \\
A-6sp &  25036 & 25075 & & 1.035 & 1.020 & & $^3$H & 15\% \\
B-d2s &  25257 & 25227 & & 0.832 & 0.995 & & $^5$K & 17\% \\
B-d2s &  25914 & 25924 & & 1.227 & 1.050 & & $^7$D & 12\% \\
A-6sp &  26071 & 26029 & & 1.304 & 1.035 & & $^5$F & 20\% \\
B-d2s &  26332 & 26345 & & 1.178 & 1.100 & & $^3$H & 6\% \\
B-d2s &  26495 & 26484 & & 1.144 & 1.035 & & $^5$G & 12\% \\
B-d2s &  26516 & 26503 & & 1.231 & 1.180 & & $^7$K & 7\% \\
B-d2s &  26580 & 26595 & & 1.079 & 1.015 & & $^3$G & 4\% \\
B-d2s &  27447 & 27474 & & 1.116 & 1.080 & & $^3$G & 7\% \\
B-ds2 &  27509 & 27524 & & 1.058 & 0.970 & & $^3$I & 15\% \\
B-d2s &  27853 & 27841 & & 1.175 & 0.990 & & $^7$F & 17\% \\
B-d2s &  28032 & 28000 & & 1.105 & 0.915 & & $^7$I & 6\% \\
B-d2s &  28146 & 28179 & & 1.053 & 0.965 & & $^3$I & 5\% \\
B-d2s &  28262 & 28280 & & 1.095 & 1.070 & & $^7$I & 7\% \\
B-d2s &  28556 & 28567 & & 1.132 & 1.090 & & $^3$G & 14\% \\
A-6sp &  28576 & 28531 & & 1.394 & 1.115 & & $^7$F & 56\% \\
B-d2s &  28689 & 28662 & & 1.136 & 0.960 & & $^3$I & 5\% \\
B-d2s &  28785 & 28723 & & 1.060 & 0.895 & & $^5$H & 4\% \\
B-d2s &  28865 & 28844 & & 1.201 & 1.100 & & $^7$D & 7\% \\
A-6sp &  28893 & 28863 & & 1.077 & 0.945 & & $^5$I & 13\% \\

%A-6sp & 15946.4 & 16028.0 & 1.000 & 0.915 & 7I 75\% \\
%B-ds2 & 17799.0 & 17790.6 & 0.895 & 1.005 & 3I 20\% \\
%B-ds2 & 19186.6 & 19226.4 & 0.981 & 0.980 & 5H 38\% \\
%B-d2s & 20988.0 & 20963.0 & 0.995 & 0.990 & 5I 27\% \\
%B-ds2 & 21319.8 & 21271.5 & 1.042 & 1.040 & 5H 9\% \\
%B-d2s & 21750.2 & 21726.7 & 1.052 & 1.000 & 5F 9\% \\
%B-d2s & 22064.7 & 22128.6 & 1.019 & 1.060 & 3I 19\% \\
%B-d2s & 22826.7 & 22736.6 & 1.065 & 1.066 & 3H 12\% \\
%B-d2s & 22972.5 & 23049.7 & 1.033 & 1.060 & 3G 11\% \\
%B-d2s & 23263.0 & 23198.1 & 1.054 & 1.085 & 7H 17\% \\
%B-d2s & 23452.2 & 23433.8 & 1.042 & 0.965 & 7H 14\% \\
%B-d2s & 23590.1 & 23573.0 & 1.083 & 1.080 & 5H 10\% \\
%B-d2s & 23819.6 & 23829.5 & 1.143 & 1.165 & 5H 13\% \\
%B-d2s & 24052.4 & 23968.3 & 0.993 & 0.980 & 5I 18\% \\
%B-d2s & 24112.9 & 24198.5 & 0.891 & 0.889 & 5K 14\% \\
%B-d2s & 24634.4 & 24586.2 & 1.098 & 1.100 & 7D 7\% \\
%B-d2s & 24743.6 & 24745.7 & 1.245 & 1.185 & 7F 16\% \\
%B-d2s & 24998.9 & 25074.7 & 1.059 & 1.020 & 3H 35\% \\
%B-d2s & 25249.0 & 25227.1 & 0.834 & 0.995 & 5K 36\% \\
%A-6sp & 26059.6 & 26029.0 & 1.327 & 1.035 & 5F 41\% \\
%B-d2s & 26258.1 & 26345.2 & 1.148 & 1.100 & 3G 12\% \\
%B-d2s & 26420.8 & 26484.1 & 1.054 & 1.035 & 7K 13\% \\
%B-d2s & 26520.4 & 26502.6 & 1.237 & 1.180 & 5F 13\% \\
%A-6sp & 26665.9 & 26595.4 & 1.102 & 1.015 & 7H 41\% \\
%B-d2s & 27435.0 & 27474.5 & 1.096 & 1.080 & 3G 13\% \\
%B-d2s & 28134.8 & 28280.5 & 1.056 & 1.070 & 3I 13\% \\
%B-d2s & 28564.2 & 28531.4 & 1.161 & 1.115 & 3G 15\% \\
%B-d2s & 28861.3 & 28843.9 & 1.103 & 1.100 & 5I 7\% \\

\bottomrule
\end{longtable}
\end{center}

%\newpage

%\captionsetup{type=table}

\begin{center}
\begin{longtable}{ l l l l l l l r r @{}}
\caption*{\textbf{Table \ref{Nd_J_6}: }Identification and comparison with NIST database values for $J = 6$ levels of odd parity configurations. All energy levels are in cm$^{-1}$.}
\label{Nd_J_6} \\
\toprule
\endfirsthead
\caption* {\textbf{Table \ref{Nd_J_6} Continued} }\\
\toprule

Conf. & \multicolumn{2}{c}{Energy } & & \multicolumn{2}{c}{Land\'e g } & & \multicolumn{2}{c}{Dom. term} \\ \cmidrule{1-9} \\
& Theory & Exp. & & Theory & Exp. & & \\
\cmidrule{2-3} \cmidrule{5-6} \\

\endhead
\cmidrule{1-9}
\endfoot
%\bottomrule
\endlastfoot
Conf. & \multicolumn{2}{c}{Energy } & & \multicolumn{2}{c}{Land\'e g } & & \multicolumn{2}{c}{Dom. term} \\ \cmidrule{1-9} \\
 & Theory & Exp. & & Theory & Exp. & & \\
 \cmidrule{2-3} \cmidrule{5-6} \\
B-d2s &  15923 & 15780 & & 0.924 & 0.945 & & $^5$K & 17\% \\
B-d2s &  17832 & 17749 & & 1.133 & 1.090 & & $^7$I & 24\% \\
A-6sp &  18184 & 18172 & & 1.236 & 1.080 & & $^7$H & 61\% \\
B-d2s &  19108 & 19152 & & 1.074 & 0.930 & & $^5$I & 8\% \\
B-d2s &  19225 & 19281 & & 1.081 & 1.055 & & $^7$H & 24\% \\
B-d2s &  19974 & 19995 & & 1.321 & 0.920 & & $^7$F & 30\% \\
B-d2s &  20518 & 20432 & & 1.101 & 1.040 & & $^5$K & 13\% \\
B-d2s &  20784 & 20703 & & 0.881 & 1.035 & & $^5$L & 17\% \\
B-ds2 &  20873 & 20839 & & 1.062 & 0.940 & & $^3$I & 15\% \\
B-d2s &  20994 & 20918 & & 0.948 & 0.840 & & $^5$L & 10\% \\
A-6sp &  21261 & 21314 & & 1.080 & 1.060 & & $^5$K & 12\% \\
B-d2s &  21695 & 21718 & & 1.066 & 0.960 & & $^5$K & 10\% \\
B-d2s &  22028 & 22050 & & 1.069 & 1.040 & & $^3$I & 9\% \\
B-d2s &  22222 & 22124 & & 1.037 & 1.170 & & $^7$I & 15\% \\
B-d2s &  22308 & 22303 & & 1.053 & 1.080 & & $^7$I & 40\% \\
B-ds2 &  22539 & 22560 & & 1.050 & 1.135 & & $^3$H & 10\% \\
B-d2s &  22759 & 22739 & & 1.000 & 0.985 & & $^5$L & 10\% \\
B-d2s &  22883 & 22871 & & 0.934 & 1.164 & & $^3$K & 12\% \\
B-d2s &  23254 & 23284 & & 1.131 & 1.040 & & $^3$I & 11\% \\
B-d2s &  23508 & 23496 & & 1.039 & 0.930 & & $^1$I & 9\% \\
B-d2s &  23649 & 23578 & & 1.089 & 0.985 & & $^5$H & 7\% \\
B-ds2 &  23718 & 23756 & & 1.105 & 1.075 & & $^3$H & 10\% \\
B-d2s &  23884 & 23889 & & 1.185 & 1.030 & & $^7$H & 26\% \\
B-d2s &  23990 & 23986 & & 1.065 & 1.100 & & $^5$I & 13\% \\
B-d2s &  24085 & 23996 & & 1.176 & 1.045 & & $^7$H & 7\% \\
A-6sp &  24558 & 24590 & & 1.007 & 1.010 & & $^3$K & 24\% \\
B-d2s &  24689 & 24702 & & 1.058 & 1.135 & & $^7$H & 13\% \\
A-6sp &  24750 & 24751 & & 0.974 & 0.925 & & $^3$K & 34\% \\
A-6sp &  24946 & 24922 & & 1.088 & 1.080 & & $^3$K & 7\% \\
B-d2s &  25033 & 24984 & & 1.170 & 1.210 & & $^3$H & 19\% \\
B-d2s &  25111 & 25115 & & 1.196 & 1.225 & & $^5$G & 17\% \\
B-d2s &  25139 & 25196 & & 1.023 & 1.120 & & $^3$K & 7\% \\
B-d2s &  25313 & 25301 & & 1.162 & 1.210 & & $^5$G & 9\% \\
B-d2s &  25526 & 25478 & & 1.084 & 1.085 & & $^5$I & 16\% \\
B-ds2 &  25581 & 25609 & & 1.064 & 0.895 & & $^5$I & 57\% \\
B-d2s &  25695 & 25662 & & 1.043 & 1.190 & & $^5$K & 10\% \\
B-d2s &  25802 & 25751 & & 1.258 & 1.200 & & $^7$F & 20\% \\
B-d2s &  26535 & 26511 & & 1.090 & 1.040 & & $^3$K & 5\% \\
B-d2s &  26727 & 26696 & & 1.129 & 1.040 & & $^5$H & 11\% \\
A-6sp &  26805 & 26816 & & 1.174 & 1.215 & & $^5$G & 6\% \\
A-6sp &  26882 & 26908 & & 1.240 & 1.125 & & $^5$G & 31\% \\

%B-d2s & 17838.5 & 17748.8 & 1.114 & 1.090 & 7I 18\% \\
%B-d2s & 19235.0 & 19281.0 & 1.138 & 1.055 & 7H 30\% \\
%B-d2s & 20521.7 & 20432.3 & 1.111 & 1.040 & 5K 20\% \\
%B-d2s & 20782.1 & 20703.1 & 0.888 & 1.035 & 5L 39\% \\
%B-d2s & 20857.9 & 20839.2 & 1.055 & 0.940 & 3I 16\% \\
%B-d2s & 21261.7 & 21314.2 & 1.128 & 1.060 & 5I 12\% \\
%B-d2s & 22024.5 & 22049.7 & 1.074 & 1.040 & 3K 8\% \\
%B-d2s & 22322.4 & 22303.0 & 1.058 & 1.080 & 7I 48\% \\
%B-d2s & 22541.3 & 22560.0 & 1.059 & 1.135 & 3H 11\% \\
%B-d2s & 22823.8 & 22738.8 & 0.993 & 0.985 & 3K 32\% \\
%B-d2s & 22992.9 & 22870.6 & 1.149 & 1.164 & 5H 43\% \\
%B-d2s & 23254.7 & 23283.6 & 1.115 & 1.040 & 3H 12\% \\
%B-d2s & 24735.6 & 24751.0 & 0.987 & 0.925 & 3K 33\% \\
%B-d2s & 24949.8 & 24922.4 & 1.103 & 1.080 & 5G 7\% \\
%B-d2s & 25030.9 & 24984.1 & 1.139 & 1.210 & 3H 22\% \\
%B-d2s & 25112.6 & 25114.8 & 1.212 & 1.225 & 5G 16\% \\
%B-d2s & 25311.8 & 25301.1 & 1.166 & 1.210 & 5G 9\% \\
%B-d2s & 25528.3 & 25478.1 & 1.081 & 1.085 & 5I 29\% \\
%B-d2s & 25802.5 & 25750.8 & 1.244 & 1.200 & 7F 18\% \\
%B-d2s & 26527.8 & 26511.0 & 1.098 & 1.040 & 3K 5\% \\
%B-d2s & 26722.6 & 26695.7 & 1.130 & 1.040 & 5H 10\% \\
%B-d2s & 26800.4 & 26816.4 & 1.183 & 1.215 & 5G 5\% \\

\bottomrule
\end{longtable}
\end{center}

%\newpage
\begin{center}
\begin{longtable}{ l l l l l l l r r @{}}
%\captionsetup{type=table}
\caption*{\textbf{Table \ref{Nd_J_7}: }Identification and comparison with NIST database values for $J = 7$ levels of odd parity configurations. All energy levels are in cm$^{-1}$.}
\label{Nd_J_7} \\
\toprule
\endfirsthead
\caption* {\textbf{Table \ref{Nd_J_7} Continued} }\\
\toprule

Conf. & \multicolumn{2}{c}{Energy } & & \multicolumn{2}{c}{Land\'e g } & & \multicolumn{2}{c}{Dom. term} \\ \cmidrule{1-9} \\
& Theory & Exp. & & Theory & Exp. & & \\
\cmidrule{2-3} \cmidrule{5-6} \\

\endhead
\cmidrule{1-9}
\endfoot
%\bottomrule
\endlastfoot
Conf. & \multicolumn{2}{c}{Energy } & & \multicolumn{2}{c}{Land\'e g } & & \multicolumn{2}{c}{Dom. term} \\ \cmidrule{1-9} \\
 & Theory & Exp. & & Theory & Exp. & & \\
 \cmidrule{2-3} \cmidrule{5-6} \\
B-d2s &  16839 & 16845 & & 1.071 & 1.120 & & $^5$K & 22\% \\
B-d2s &  17312 & 17290 & & 1.188 & 1.070 & & $^7$H & 29\% \\
B-d2s &  18121 & 18257 & & 0.852 & 0.955 & & $^5$M & 42\% \\
B-d2s &  19682 & 19746 & & 1.279 & 1.090 & & $^7$G & 21\% \\
B-ds2 &  21090 & 21026 & & 1.160 & 1.235 & & $^5$H & 39\% \\
B-d2s &  21306 & 21286 & & 1.035 & 1.050 & & $^5$K & 22\% \\
B-d2s &  21365 & 21412 & & 1.064 & 1.067 & & $^5$K & 12\% \\
A-6sp &  21910 & 21909 & & 1.086 & 0.970 & & $^5$K & 14\% \\
A-6sp &  22055 & 22042 & & 1.376 & 1.020 & & $^7$G & 79\% \\
B-d2s &  22334 & 22320 & & 1.133 & 1.128 & & $^5$I & 15\% \\
A-6sp &  22513 & 22483 & & 1.053 & 1.120 & & $^5$K & 23\% \\ 
B-d2s &  23084 & 22939 & & 1.059 & 1.065 & & $^7$I & 12\% \\
A-6sp &  23518 & 23518 & & 1.147 & 1.052 & & $^5$I & 16\% \\
B-d2s &  23779 & 23745 & & 1.097 & 1.040 & & $^5$L & 13\% \\
B-d2s &  24072 & 23991 & & 1.072 & 1.225 & & $^5$K & 10\% \\
A-6sp &  24177 & 24213 & & 1.255 & 1.060 & & $^5$H & 36\% \\
B-d2s &  24676 & 24730 & & 1.216 & 1.150 & & $^5$H & 19\% \\
B-d2s &  24825 & 24968 & & 1.120 & 1.145 & & $^5$H & 8\% \\
B-d2s &  25079 & 25064 & & 1.136 & 1.100 & & $^3$I & 21\% \\
B-d2s &  25164 & 25198 & & 1.222 & 1.120 & & $^5$H & 17\% \\
B-d2s &  25485 & 25504 & & 1.140 & 1.115 & & $^5$I & 7\% \\
B-d2s &  25569 & 25596 & & 1.234 & 1.095 & & $^7$G & 17\% \\
B-ds2 &  25919 & 25849 & & 0.951 & 1.085 & & $^3$L & 57\% \\
B-d2s &  25954 & 25886 & & 1.068 & 0.965 & & $^1$K & 11\% \\
B-ds2 &  26162 & 26155 & & 1.116 & 0.950 & & $^5$I & 11\% \\
A-6sp &  26270 & 26327 & & 1.124 & 1.250 & & $^3$K & 14\% \\
B-d2s &  26424 & 26395 & & 1.118 & 1.080 & & $^3$I & 8\% \\
B-d2s &  26509 & 26635 & & 1.060 & 1.125 & & $^3$L & 13\% \\

%B-d2s & 16847.4 & 16845.3 & 1.072 & 1.120 & 5K 44\% \\
%B-d2s & 17313.5 & 17289.6 & 1.202 & 1.070 & 7H 31\% \\
%B-d2s & 18108.4 & 18256.8 & 0.856 & 0.955 & 5M 40\% \\
%B-d2s & 19610.9 & 19746.2 & 1.058 & 1.090 & 5I 11\% \\
%B-ds2 & 21105.4 & 21025.5 & 1.160 & 1.235 & 5H 39\% \\
%B-d2s & 21315.0 & 21285.9 & 1.035 & 1.050 & 5K 21\% \\
%B-d2s & 21365.6 & 21411.5 & 1.057 & 1.067 & 5K 18\% \\
%B-d2s & 22325.6 & 22320.2 & 1.130 & 1.128 & 5I 21\% \\
%B-d2s & 22510.8 & 22482.7 & 1.055 & 1.120 & 5K 7\% \\
%B-d2s & 23439.8 & 23517.7 & 1.133 & 1.052 & 7I 20\% \\
%B-d2s & 23788.0 & 23876.9 & 1.103 & 1.075 & 5L 20\% \\
%B-d2s & 24075.4 & 24212.9 & 1.091 & 1.060 & 5K 9\% \\
%B-d2s & 24688.4 & 24729.5 & 1.214 & 1.150 & 5H 18\% \\
%B-d2s & 24803.3 & 24968.4 & 1.108 & 1.145 & 3K 17\% \\
%B-d2s & 25078.0 & 25063.7 & 1.153 & 1.100 & 3I 21\% \\
%B-d2s & 25162.4 & 25197.7 & 1.220 & 1.120 & 5H 15\% \\
%B-d2s & 25482.8 & 25503.9 & 1.142 & 1.115 & 5I 12\% \\

\bottomrule
\end{longtable}
\end{center}

%\newpage
%\captionsetup{type=table}
\begin{center}
\begin{longtable}{ l l l l l l l r r @{}}
\caption*{\textbf{Table \ref{Nd_J_8}: }Identification and comparison with NIST database values for $J = 8$ levels of odd parity configurations. All energy levels are in cm$^{-1}$.}
\label{Nd_J_8} \\

\toprule
\endfirsthead
\caption* {\textbf{Table \ref{Nd_J_8} Continued} }\\
\toprule

Conf. & \multicolumn{2}{c}{Energy } & & \multicolumn{2}{c}{Land\'e g } & & \multicolumn{2}{c}{Dom. term} \\ \cmidrule{1-9} \\
& Theory & Exp. & & Theory & Exp. & & \\
\cmidrule{2-3} \cmidrule{5-6} \\

\endhead
\cmidrule{1-9}
\endfoot
%\bottomrule
\endlastfoot
Conf. & \multicolumn{2}{c}{Energy } & & \multicolumn{2}{c}{Land\'e g } & & \multicolumn{2}{c}{Dom. term} \\ \cmidrule{1-9} \\
 & Theory & Exp. & & Theory & Exp. & & \\
 \cmidrule{2-3} \cmidrule{5-6} \\
B-d2s &  18080 & 17973 & & 1.141 & 1.200 & & $^7$K & 38\% \\
A-6sp &  19921 & 19862 & & 1.162 & 1.290 & & $^7$H & 14\% \\
A-6sp &  23542 & 23474 & & 1.135 & 1.123 & & $^5$K & 19\% \\ 
B-d2s &  23665 & 23653 & & 1.206 & 1.128 & & $^7$I & 27\% \\ 
A-6sp &  23972 & 24078 & & 1.167 & 1.070 & & $^5$K & 14\% \\
B-d2s &  24651 & 24688 & & 1.183 & 1.155 & & $^7$I & 17\% \\ 
B-d2s &  24756 & 24773 & & 1.180 & 1.145 & & $^5$I & 21\% \\ 
B-d2s &  25191 & 25191 & & 1.191 & 1.195 & & $^5$I & 16\% \\ 
B-d2s &  25212 & 25281 & & 1.174 & 1.115 & & $^5$I & 15\% \\ 
B-d2s &  25477 & 25383 & & 1.157 & 1.140 & & $^5$I & 17\% \\ 
B-d2s &  25538 & 25514 & & 1.100 & 1.140 & & $^3$L & 12\% \\ 
B-d2s &  26738 & 26783 & & 1.087 & 1.118 & & $^3$L & 15\% \\
B-ds2 &  26996 & 27044 & & 1.026 & 1.015 & & $^3$L & 53\% \\
B-d2s &  27168 & 27131 & & 1.196 & 1.170 & & $^5$I & 13\% \\

%B-d2s & 18115.8 & 17973.3 & 1.143 & 1.200 & 7K 40\% \\
%B-d2s & 19917.0 & 19862.5 & 1.176 & 1.290 & 5I 30\% \\
%B-d2s & 23536.0 & 23474.0 & 1.128 & 1.123 & 3K 6\% \\
%B-d2s & 23670.9 & 23652.7 & 1.207 & 1.128 & 7I 26\% \\
%B-d2s & 24668.6 & 24688.2 & 1.176 & 1.155 & 7I 19\% \\
%B-d2s & 24796.0 & 24773.3 & 1.189 & 1.145 & 5I 47\% \\
%B-d2s & 25220.0 & 25190.7 & 1.168 & 1.195 & 5I 16\% \\
%B-d2s & 25544.1 & 25513.7 & 1.169 & 1.140 & 5I 27\% \\

\bottomrule
\end{longtable}
\end{center}

%\newpage
%\captionsetup{type=table}
\begin{center}
\begin{longtable}{ l l l l l l l r r @{}}
\caption*{\textbf{Table \ref{Nd_J_9}: }Identification and comparison with NIST database values for $J = 9$ levels of odd parity configurations. All energy levels are in cm$^{-1}$.}
\label{Nd_J_9} \\
\toprule
\endfirsthead
\caption* {\textbf{Table \ref{Nd_J_4} Continued} }\\
\toprule

Conf. & \multicolumn{2}{c}{Energy } & & \multicolumn{2}{c}{Land\'e g } & & \multicolumn{2}{c}{Dom. term} \\ \cmidrule{1-9} \\
& Theory & Exp. & & Theory & Exp. & & \\
\cmidrule{2-3} \cmidrule{5-6} \\

\endhead
\cmidrule{1-9}
\endfoot
%\bottomrule
\endlastfoot
Conf. & \multicolumn{2}{c}{Energy } & & \multicolumn{2}{c}{Land\'e g } & & \multicolumn{2}{c}{Dom. term} \\ \cmidrule{1-9} \\
 & Theory & Exp. & & Theory & Exp. & & \\
 \cmidrule{2-3} \cmidrule{5-6} \\
B-d2s &  24933 & 24935 & & 1.098 & 1.150 & & $^3$M & 20\% \\ 
B-d2s &  25224 & 25142 & & 1.102 & 1.120 & & $^3$M & 17\% \\ 
B-d2s &  26424 & 26511 & & 1.121 & 1.165 & & $^3$L & 28\% \\ 
B-d2s &  27862 & 27842 & & 1.121 & 1.160 & & $^3$L & 24\% \\
B-ds2 &  28734 & 28781 & & 1.105 & 1.215 & & $^3$L & 72\% \\ 
B-d2s &  28983 & 29061 & & 1.220 & 1.015 & & $^5$K & 25\% \\

%B-d2s & 24939.5 & 24935.0 & 1.090 & 1.150 & 3M 30\% \\
%B-d2s & 25218.9 & 25141.5 & 1.103 & 1.120 & 3M 31\% \\
%B-d2s & 26440.4 & 26510.9 & 1.121 & 1.165 & 3L 36\% \\

\bottomrule

\end{longtable}
\end{center}